\documentclass{article}%
\usepackage{amsfonts}
\usepackage{amssymb}
\usepackage{amsmath}
\usepackage{graphicx}%
\providecommand{\U}[1]{\protect\rule{.1in}{.1in}}
\newtheorem{corollary}{Corollary}

\newtheorem{proposition}{Proposition}

\begin{document}

\title{What can (partition) logic contribute to information theory?}
\author{David Ellerman\\University of California at Riverside}
\maketitle

\begin{abstract}
\noindent Logical probability theory was developed as a quantitative measure
based on Boole's logic of subsets. But information theory was developed into a
mature theory by Claude Shannon with no such connection to logic. A recent
development in logic changes this situation. In category theory, the notion of
a subset is dual to the notion of a quotient set or partition, and recently
the logic of partitions has been developed in a parallel relationship to the
Boolean logic of subsets (subset logic is usually mis-specified as the special
case of propositional logic). What then is the quantitative measure based on
partition logic in the same sense that logical probability theory is based on
subset logic? It is a measure of information that is named "logical entropy"
in view of that logical basis. This paper develops the notion of logical
entropy and the basic notions of the resulting logical information theory.
Then an extensive comparison is made with the corresponding notions based on
Shannon entropy.


\end{abstract}
\tableofcontents

\section{Introduction}

This paper develops the application of the logic of partitions
\cite{ell:intropartlogic} to information theory. Partitions are dual (in a
category-theoretic sense) to subsets. George Boole developed the notion of
logical probability \cite{boole:lot} as the normalized counting measure on
subsets in his logic of subsets. This paper develops the normalized counting
measure on partitions as the analogous quantitative treatment in the logic of
partitions. The resulting measure is a new logical derivation of an old
formula measuring diversity and distinctions, e.g., the Gini-Simpson index of
diversity, that goes back to the early 20th century \cite{gini:vem}. In view
of the idea of information as being based on distinctions (see next section),
I refer to this logical measure of distinctions as "logical entropy".

This raises the question of the relationship of logical entropy to the
standard notion of Shannon entropy. Firstly, logical entropy directly counts
the distinctions (as defined in partition logic) whereas Shannon entropy, in
effect, counts the minimum number of binary partitions (or yes/no questions)
it takes, on average, to uniquely determine or designate the distinct
entities. Since that gives a binary code for the distinct entities, the
Shannon theory (unlike the logical theory) is perfectly adapted for the theory
of coding and communications.

The second way to relate the logical theory and the Shannon theory is to
consider the relationship between the compound notions (e.g., conditional
entropy, joint entropy, and mutual information) in the two theories. Logical
entropy is a measure in the mathematical sense, so as with any measure, the
compound formulas satisfy the usual Venn-diagram relationships. The compound
notions of Shannon entropy are \textit{defined} so that they also satisfy
similar Venn diagram relationships. However, as various information theorists,
principally Lorne Campbell, have noted \cite{camp:meas}, Shannon entropy is
not a measure (outside of the special case of $2^{n}$ equiprobable distinct
entities where it is the count $n$ of the number of yes/no questions necessary
to unique determine the distinct entities)--so one can conclude only that the
"analogies provide a convenient mnemonic" \cite[p. 112]{camp:meas} in terms of
the usual Venn diagrams for measures. Campbell wondered if there might be a
"deeper foundation" \cite[p. 112]{camp:meas} to clarify how the Shannon
formulas can defined to satisfy the measure-like relations in spite of not
being a measure. That question is addressed in this paper by showing that
there is a transformation of formulas that transforms each of the logical
entropy compound formulas into the corresponding Shannon entropy compound
formula, and the transform preserves the Venn diagram relationships that
automatically hold for measures. This "dit-bit transform" is heuristically
motivated by showing how certain counts of distinctions ("dits") can be
converted in counts of binary partitions ("bits").

Moreover, Campbell remarked that it would be "particularly interesting" and
"quite significant" if there was an entropy measure of sets so that joint
entropy corresponded to the measure of the union of sets, conditional entropy
to the difference of sets, and mutual information to the intersection of sets
\cite[p. 113]{camp:meas}. Logical entropy precisely satisfies those
requirements, so we turn to the underlying idea of information as a measure of distinctions.

\section{Logical information as the measure of distinctions}

There is now a widespread view that information is fundamentally about
differences, distinguishability, and distinctions. As Charles H. Bennett, one
of the founders of quantum information theory, put it:

\begin{quotation}
So information really is a very useful abstraction. It is the notion of
distinguishability abstracted away from what we are distinguishing, or from
the carrier of information. \cite[p. 155]{bennett:qinfo}
\end{quotation}

\noindent This view even has an interesting history. In James Gleick's book,
\textit{The Information: A History, A Theory, A Flood}, he noted the focus on
differences in the seventeenth century polymath, John Wilkins, who was a
founder of the Royal Society. In $1641$, the year before Newton was born,
Wilkins published one of the earliest books on cryptography, \textit{Mercury
or the Secret and Swift Messenger}, which not only pointed out the fundamental
role of differences but noted that any (finite) set of different things could
be encoded by words in a binary code.

\begin{quotation}
\noindent For in the general we must note, That whatever is capable of a
competent Difference, perceptible to any Sense, may be a sufficient Means
whereby to express the Cogitations. It is more convenient, indeed, that these
Differences should be of as great Variety as the Letters of the Alphabet; but
it is sufficient if they be but twofold, because Two alone may, with somewhat
more Labour and Time, be well enough contrived to express all the rest.
\cite[Chap. XVII, p. 69]{wilkins:merc}
\end{quotation}

\noindent Wilkins explains that a five letter binary code would be sufficient
to code the letters of the alphabet since $2^{5}=32$.

\begin{quotation}
\noindent Thus any two Letters or Numbers, suppose $A.B$. being transposed
through five Places, will yield Thirty Two Differences, and so consequently
will superabundantly serve for the Four and twenty Letters... .\cite[Chap.
XVII, p. 69]{wilkins:merc}
\end{quotation}

\noindent As Gleick noted:

\begin{quotation}
\noindent Any difference meant a binary choice. Any binary choice began the
expressing of cogitations. Here, in this arcane and anonymous treatise of
$1641$, the essential idea of information theory poked to the surface of human
thought, saw its shadow, and disappeared again for [three] hundred years.
\cite[p. 161]{gleick:info} \noindent
\end{quotation}

\noindent Thus \textit{counting distinctions} \cite{ell:countingdits} would
seem the right way to measure information,\footnote{This paper is about what
Adriaans and Benthem call "Information B: Probabilistic,
information-theoretic, measured quantitatively", not about "Information A:
knowledge, logic, what is conveyed in informative answers" where the
connection to philosophy and logic is built-in from the beginning. Likewise,
the paper is not about Kolmogorov-style "Information C: Algorithmic, code
compression, measured quantitatively." \cite[p. 11]{adriaans-benthem:philinfo}%
} and that is the measure that emerges naturally out of partition logic--just
as finite logical probability emerges naturally as the measure of
\textit{counting elements} in Boole's subset logic.

Although usually named after the special case of `propositional' logic, the
general case is Boole's logic of subsets of a universe $U$ (the special case
of $U=1$ allows the propositional interpretation since the only subsets are
$1$ and $\emptyset$ standing for truth and falsity). Category theory shows
that is a duality between sub-sets and quotient-sets (= partitions =
equivalence relations), and that allowed the recent development of the dual
logic of partitions (\cite{ell:partitions}, \cite{ell:intropartlogic}). As
indicated in the title of his book, \textit{An Investigation of the Laws of
Thought on which are founded the Mathematical Theories of Logic and
Probabilities} \cite{boole:lot}, Boole also developed the normalized counting
measure on subsets of a finite universe $U$ which was finite logical
probability theory. When the same mathematical notion of the normalized
counting measure is applied to the partitions on a finite universe set $U$
(when the partition is represented as the complement of the corresponding
equivalence relation on $U\times U$) then the result is the formula for
logical entropy.

In addition to the philosophy of information literature
\cite{adriaans-benthem:philinfo}, there is a whole sub-industry in mathematics
concerned with different notions of `entropy' or `information'
(\cite{aczel-d:measures}; see \cite{tsallis:nonext-sm} for a recent
`extensive' analysis) that is long on formulas and `intuitive axioms' but
short on interpretations. Out of that plethora of definitions, logical entropy
is the \textit{measure} (in the technical sense of measure) of information
that arises out of partition logic just as logical probability theory arises
out of subset logic.

The logical notion of information-as-distinctions supports the view that the
notion of information is a more primative notion than probability and should
be based on finite combinatorics. As Kolmogorov put it:

\begin{quote}
\noindent Information theory must precede probability theory, and not be based
on it. By the very essence of this discipline, the foundations of information
theory have a finite combinatorial character. \cite[p. 39]{kolmogor:combfound}
\end{quote}

\noindent Logical information theory starts simply with a set of distinctions
defined by a partition on $U$, where a distinction is an ordered pair of
elements of $U$ in distinct blocks of the partition. Thus the \textit{set of
distinctions} ("ditset") or \textit{information set} ("infoset") associated
with the partition is just the complement of the equivalence relation
associated with the partition. To get a quantitative measure of information,
any probability distribution on $U$ defines a product probability measure on
$U\times U$, and the logical entropy is simply that probability measure of the
information set. In this manner, the logical theory of
information-as-distinctions starts with the information set (set of
distinctions) as a finite combinatorial object and then for any probability
measure on the underlying set, the product probability measure on the
information set gives the quantitative notion of logical entropy.

\section{Duality of subsets and partitions}

Logical entropy is to the logic of partitions as logical probability is to the
Boolean logic of subsets. Hence we will start with a brief review of the
relationship between these two dual forms of mathematical logic.

Modern category theory shows that the concept of a subset dualizes to the
concept of a quotient set, equivalence relation, or partition. F. William
Lawvere called a subset or, in general, a subobject a \textquotedblleft
part\textquotedblright\ and then noted: \textquotedblleft The dual notion
(obtained by reversing the arrows) of `part' is the notion of
\textit{partition}.\textquotedblright\ \cite[p. 85]{law:sfm} That suggests
that the Boolean logic of subsets (usually named after the special case of
propositions as `propositional' logic) should have a dual logic of partitions
(\cite{ell:partitions}, \cite{ell:intropartlogic}).

A \textit{partition} $\pi=\left\{  B_{1},...,B_{m}\right\}  $ on $U$ is a set
of subsets, called cells or blocks, $B_{i}$ that are mutually disjoint and
jointly exhaustive ($\cup_{i}B_{i}=U$). In the duality between subset logic
and partition logic, the dual to the notion of an `element' of a subset is the
notion of a `distinction' of a partition, where $\left(  u,u^{\prime}\right)
\in U\times U$ is a \textit{distinction} or \textit{dit} of $\pi$ if the two
elements are in different blocks, i.e., the `dits' of a partition are dual to
the `its' (or elements) of a subset. Let $\operatorname*{dit}\left(
\pi\right)  \subseteq U\times U$ be the set of distinctions or \textit{ditset}
of $\pi$. Thus the \textit{information set} or \textit{infoset} associated
with a partition $\pi$ is ditset $\operatorname*{dit}\left(  \pi\right)  $.
Similarly an \textit{indistinction} or \textit{indit} of $\pi$ is a pair
$\left(  u,u^{\prime}\right)  \in U\times U$ in the same block of $\pi$. Let
$\operatorname*{indit}\left(  \pi\right)  \subseteq U\times U$ be the set of
indistinctions or \textit{inditset }of $\pi$. Then $\operatorname*{indit}%
\left(  \pi\right)  $ is the equivalence relation associated with $\pi$ and
$\operatorname*{dit}\left(  \pi\right)  =U\times U-\operatorname*{indit}%
\left(  \pi\right)  $ is the complementary binary relation that might be
called a\textit{ partition relation} or an \textit{apartness relation}.

\section{Classical subset logic and partition logic}

The algebra associated with the subsets $S\subseteq U$ is, of course, the
Boolean algebra $\wp\left(  U\right)  $ of subsets of $U$ with the partial
order as the inclusion of elements. The corresponding algebra of partitions
$\pi$ on $U$ is the \textit{partition algebra} $\prod\left(  U\right)  $
defined as follows:

\begin{itemize}
\item the \textit{partial order} $\sigma\preceq\pi$ of partitions
$\sigma=\left\{  C,C^{\prime},...\right\}  $ and $\pi=\left\{  B,B^{\prime
},...\right\}  $ holds when $\pi$ \textit{refines} $\sigma$ in the sense that
for every block $B\in\pi$ there is a block $C\in\sigma$ such that $B\subseteq
C$, or, equivalently, using the element-distinction pairing, the partial order
is the inclusion of distinctions: $\sigma\preceq\pi$ if and only if (iff)
$\operatorname*{dit}\left(  \sigma\right)  \subseteq\operatorname*{dit}\left(
\pi\right)  $;

\item the minimum or bottom partition is the \textit{indiscrete partition} (or
blob) $\mathbf{0}=\left\{  U\right\}  $ with one block consisting of all of
$U$;

\item the maximum or top partition is the \textit{discrete partition}
$\mathbf{1}=\left\{  \left\{  u_{j}\right\}  \right\}  _{j=1,...,n}$
consisting of singleton blocks;

\item the \textit{join} $\pi\vee\sigma$ is the partition whose blocks are the
non-empty intersections $B\cap C$ of blocks of $\pi$ and blocks of $\sigma$,
or, equivalently, using the element-distinction pairing, $\operatorname*{dit}%
\left(  \pi\vee\sigma\right)  =\operatorname*{dit}\left(  \pi\right)
\cup\operatorname*{dit}\left(  \sigma\right)  $;

\item the \textit{meet} $\pi\wedge\sigma$ is the partition whose blocks are
the equivalence classes for the equivalence relation generated by: $u_{j}\sim
u_{j^{\prime}}$ if $u_{j}\in B\in\pi$, $u_{j^{\prime}}\in C\in\sigma$, and
$B\cap C\neq\emptyset$; and

\item $\sigma\Rightarrow\pi$ is the \textit{implication partition} whose
blocks are: (1) the singletons $\left\{  u_{j}\right\}  $ for $u_{j}\in
B\in\pi$ if there is a $C\in\sigma$ such that $B\subseteq C$, or (2) just
$B\in\pi$ if there is no $C\in\sigma$ with $B\subseteq C$, so that trivially:
$\sigma\Rightarrow\pi=\mathbf{1}$ iff $\sigma\preceq\pi$.\footnote{There is a
general method to define operations on partitions corresponding to operations
on subsets (\cite{ell:partitions}, \cite{ell:intropartlogic}) but the lattice
operations of join and meet, and the implication operation are sufficient to
define a partition algebra $\prod\left(  U\right)  $ parallel to the familiar
powerset Boolean algebra $\wp\left(  U\right)  $.}
\end{itemize}

The logical partition operations can also be defined in terms of the
corresponding logical operations on subsets. A ditset $\operatorname*{dit}%
\left(  \pi\right)  $ of a partition on $U$ is a subset of $U\times U$ of a
particular kind, namely the complement of an equivalence relation. An
\textit{equivalence relation} is reflexive, symmetric, and transitive. Hence
the complement, i.e., a partition relation (or apartness relation), is a
subset $P\subseteq U\times U$ that is:

\begin{enumerate}
\item irreflexive (or anti-reflexive), $P\cap\Delta=\emptyset$ (where
$\Delta=\left\{  \left(  u,u\right)  :u\in U\right\}  $ is the
\textit{diagonal});

\item symmetric, $\left(  u,u^{\prime}\right)  \in P$ implies $\left(
u^{\prime},u\right)  \in P$; and

\item anti-transitive (or co-transitive), if $\left(  u,u^{\prime\prime
}\right)  \in P$ then for any $u^{\prime}\in U$, $\left(  u,u^{\prime}\right)
\in P$ or $\left(  u^{\prime},u^{\prime\prime}\right)  \in P$.
\end{enumerate}

Given any subset $S\subseteq U\times U$, the
\textit{reflexive-symmetric-transitive (rst) closure} $\overline{S^{c}}$ of
the complement $S^{c}$ is the smallest equivalence relation containing $S^{c}%
$, so its complement is the largest partition relation contained in $S$, which
is called the \textit{interior} $\operatorname*{int}\left(  S\right)  $ of
$S$. This usage is consistent with calling the subsets that equal their
rst-closures \textit{closed subsets} of $U\times U$ (so closed subsets =
equivalence relations) so the complements are the \textit{open subsets} (=
partition relations). However it should be noted that the rst-closure is not a
topological closure since the closure of a union is not necessarily the union
of the closures, so the `open' subsets do not form a topology on $U\times U$.
Indeed, any two nonempty open sets have a nonempty intersection.

The interior operation $\operatorname*{int}:\wp\left(  U\times U\right)
\rightarrow\wp\left(  U\times U\right)  $ provides a universal way to define
logical operations on partitions from the corresponding logical subset
operations in Boolean logic:

\begin{center}
apply the subset operation to the ditsets and then, if necessary,

take the interior to obtain the ditset of the partition operation.
\end{center}

Since the same operations can be defined for subsets and partitions, one can
interpret a formula $\Phi\left(  \pi,\sigma,...\right)  $ either way as a
subset or a partition. Given either subsets on or partitions of $U$
substituted for the variables $\pi$, $\sigma$,..., one can apply,
respectively, subset or partition operations to evaluate the whole formula.
Since $\Phi\left(  \pi,\sigma,...\right)  $ is either a subset or a partition,
the corresponding proposition is \textquotedblleft$u$ is an element of
$\Phi\left(  \pi,\sigma,...\right)  $\textquotedblright\ or \textquotedblleft%
$\left(  u,u^{\prime}\right)  $ is a distinction of $\Phi\left(  \pi
,\sigma,...\right)  $\textquotedblright. And then the definitions of a valid
formula are also parallel, namely, no matter what is substituted for the
variables, the whole formula evaluates to the top of the algebra. In that
case, the subset $\Phi\left(  \pi,\sigma,...\right)  $ contains all elements
of $U$, i.e., $\Phi\left(  \pi,\sigma,...\right)  =U$, or the partition
$\Phi\left(  \pi,\sigma,...\right)  $ distinguishes all pairs $\left(
u,u^{\prime}\right)  $ for distinct elements of $U$, i.e., $\Phi\left(
\pi,\sigma,...\right)  =\mathbf{1}$. The parallelism between the dual logics
is summarized in the following table 1.

\begin{center}%
\begin{tabular}
[c]{|c||c|c|}\hline
Table 1 & Subset logic & Partition logic\\\hline\hline
`Elements' (its or dits) & Elements $u$ of $S$ & Dits $\left(  u,u^{\prime
}\right)  $ of $\pi$\\\hline
Inclusion of `elements' & Inclusion $S\subseteq T$ & Refinement:
$\operatorname*{dit}\left(  \sigma\right)  \subseteq\operatorname*{dit}\left(
\pi\right)  $\\\hline
Top of order = all `elements' & $U$ all elements & $\operatorname*{dit}%
(\mathbf{1)}=U^{2}-\Delta$, all dits\\\hline
Bottom of order = no `elements' & $\emptyset$ no elements &
$\operatorname*{dit}(\mathbf{0)=\emptyset}$, no dits\\\hline
Variables in formulas & Subsets $S$ of $U$ & Partitions $\pi$ on $U$\\\hline
Operations: $\vee,\wedge,\Rightarrow,...$ & Subset ops. & Partition
ops.\\\hline
Formula $\Phi(x,y,...)$ holds & $u$ element of $\Phi(S,T,...)$ & $\left(
u,u^{\prime}\right)  $ dit of $\Phi(\pi,\sigma,...)$\\\hline
Valid formula & $\Phi(S,T,...)=U$, $\forall S,T,...$ & $\Phi(\pi
,\sigma,...)=\mathbf{1}$, $\forall\pi,\sigma,...$\\\hline
\end{tabular}

Table 1: Duality between subset logic and partition logic
\end{center}

\section{Classical logical probability and logical entropy}

George Boole \cite{boole:lot} extended his logic of subsets to finite logical
probability theory where, in the equiprobable case, the probability of a
subset $S$ (event) of a finite universe set (outcome set or sample space)
$U=\left\{  u_{1},...,u_{n}\right\}  $ was the number of elements in $S$ over
the total number of elements: $\Pr\left(  S\right)  =\frac{\left\vert
S\right\vert }{\left\vert U\right\vert }=\sum_{u_{j}\in S}\frac{1}{\left\vert
U\right\vert }$. Laplace's classical finite probability theory
\cite{laplace:probs} also dealt with the case where the outcomes were assigned
real point probabilities $p=\left\{  p_{1},...,p_{n}\right\}  $ (where
$p_{j}\geq0$ and $\sum_{j}p_{j}=1$) so rather than summing the equal
probabilities $\frac{1}{\left\vert U\right\vert }$, the point probabilities of
the elements were summed: $\Pr\left(  S\right)  =\sum_{u_{j}\in S}%
p_{j}=p\left(  S\right)  $--where the equiprobable formula is for $p_{j}%
=\frac{1}{\left\vert U\right\vert }$ for $j=1,...,n$. The conditional
probability of an event $T\subseteq U$ given an event $S$ is $\Pr\left(
T|S\right)  =\frac{p\left(  T\cap S\right)  }{p\left(  S\right)  }$.

Then we may mimic Boole's move going from the logic of subsets to the finite
logical probabilities of subsets by starting with the logic of partitions and
using the dual relation between elements and distinctions. The dual notion to
probability turns out to be `information content' or `entropy' so we define
the \textit{logical entropy} of $\pi=\left\{  B_{1,}...,B_{m}\right\}  $,
denoted $h\left(  \pi\right)  $, as the size of the ditset
$\operatorname*{dit}\left(  \pi\right)  \subseteq U\times U$ normalized by the
size of $U\times U$:

\begin{center}
$h\left(  \pi\right)  =\frac{\left\vert \operatorname*{dit}\left(  \pi\right)
\right\vert }{\left\vert U\times U\right\vert }=\sum_{\left(  u_{j}%
,u_{k}\right)  \in\operatorname*{dit}\left(  \pi\right)  }\frac{1}{\left\vert
U\right\vert }\frac{1}{\left\vert U\right\vert }$

Logical entropy of $\pi$ (equiprobable case).
\end{center}

\noindent This is just the product probability measure of the equiprobable or
uniform probability distribution on $U$ applied to the information set or
ditset $\operatorname*{dit}\left(  \pi\right)  $. The inditset of $\pi$ is
$\operatorname*{indit}\left(  \pi\right)  =\cup_{i=1}^{m}\left(  B_{i}\times
B_{i}\right)  $ so where $p\left(  B_{i}\right)  =\frac{|B_{i}|}{\left\vert
U\right\vert }$ in the equiprobable case, we have:

\begin{center}
$h\left(  \pi\right)  =\frac{\left\vert \operatorname*{dit}\left(  \pi\right)
\right\vert }{\left\vert U\times U\right\vert }=\frac{\left\vert U\times
U\right\vert -\sum_{i=1}^{m}\left\vert B_{i}\times B_{i}\right\vert
}{\left\vert U\times U\right\vert }=1-\sum_{i=1}^{m}\left(  \frac{\left\vert
B_{i}\right\vert }{\left\vert U\right\vert }\right)  ^{2}=1-\sum_{i=1}%
^{m}p\left(  B_{i}\right)  ^{2}$.
\end{center}

The corresponding definition for the case of point probabilities $p=\left\{
p_{1},...,p_{n}\right\}  $ is to just add up the probabilities of getting a
particular distinction:

\begin{center}
$h_{p}\left(  \pi\right)  =\sum_{\left(  u_{j},u_{k}\right)  \in
\operatorname*{dit}\left(  \pi\right)  }p_{j}p_{k}$

Logical entropy of $\pi$ with point probabilities $p$.
\end{center}

\noindent Taking $p\left(  B_{i}\right)  =\sum_{u_{j}\in B_{i}}p_{j}$, the
logical entropy with point probabilities is:

\begin{center}
$h_{p}\left(  \pi\right)  =\sum_{\left(  u_{j},u_{k}\right)  \in
\operatorname*{dit}\left(  \pi\right)  }p_{j}p_{k}=\sum_{i\neq i^{\prime}%
}p\left(  B_{i}\right)  p\left(  B_{i^{\prime}}\right)  =2\sum_{i<i^{\prime}%
}p\left(  B_{i}\right)  p\left(  B_{i^{\prime}}\right)  =1-\sum_{i=1}%
^{m}p\left(  B_{i}\right)  ^{2}$.
\end{center}

Instead of being given a partition $\pi=\left\{  B_{1},...,B_{m}\right\}  $ on
$U$ with point probabilities $p_{j}$ defining the finite probability
distribution of block probabilities $\left\{  p\left(  B_{i}\right)  \right\}
_{i}$, one might be given only a finite probability distribution $p=\left\{
p_{1},...,p_{m}\right\}  $. Then substituting $p_{i}$ for $p\left(
B_{i}\right)  $ gives the:

\begin{center}
$h\left(  p\right)  =1-\sum_{i=1}^{m}p_{i}^{2}=\sum_{i\neq j}p_{i}p_{j}$

Logical entropy of a finite probability distribution.
\end{center}

\noindent Since $1=\left(  \sum_{i=1}^{n}p_{i}\right)  ^{2}=\sum_{i}p_{i}%
^{2}+\sum_{i\neq j}p_{i}p_{j}$, we again have the logical entropy $h\left(
p\right)  $ as the probability $\sum_{i\neq j}p_{i}p_{j}$ of drawing a
distinction in two independent samplings of the probability distribution $p$.

That two-draw probability interpretation follows from the important fact that
logical entropy is always the value of a probability measure. The product
probability measure on the subsets $S\subseteq U\times U$ is:

\begin{center}
$\mu\left(  S\right)  =\sum\left\{  p_{i}p_{j}:\left(  u_{i},u_{j}\right)  \in
S\right\}  $

\textit{Product measure} on $U\times U$.
\end{center}

\noindent Then the logical entropy $h\left(  p\right)  =\mu\left(
\operatorname*{dit}(\mathbf{1}_{U})\right)  $ is just the product measure of
the information set or ditset $\operatorname*{dit}\left(  \mathbf{1}%
_{U}\right)  =U\times U-\Delta$ of the discrete partition $\mathbf{1}_{U}$ on
$U$. 

There are also parallel \textquotedblleft element $\leftrightarrow$
distinction\textquotedblright\ probabilistic interpretations:

\begin{itemize}
\item $\Pr\left(  S\right)  =p_{S}$ is the probability that a single draw,
sample, or experiment with $U$ gives a \textit{element} $u_{j}\ $of $S$, and

\item $h_{p}\left(  \pi\right)  =\mu\left(  \operatorname*{dit}\left(
\pi\right)  \right)  =\sum_{\left(  u_{j},u_{k}\right)  \in\operatorname*{dit}%
\left(  \pi\right)  }p_{j}p_{k}=\sum_{i\neq i^{\prime}}p\left(  B_{i}\right)
p\left(  B_{i^{\prime}}\right)  =1-\sum_{i}P\left(  B_{i}\right)  ^{2}$ is the
probability that two independent (with replacement) draws, samples, or
experiments with $U$ gives a \textit{distinction} $\left(  u_{j},u_{k}\right)
$ of $\pi$, or if we interpret the independent experiments as sampling from
the set of blocks $\pi=\left\{  B_{i}\right\}  $, then it is the probability
of getting distinct blocks.
\end{itemize}

\noindent In probability theory, when a random draw gives an outcome $u_{j}$
in the subset or event $S$, we say the event $S$ \textit{occurs}, and in
logical information theory, when the random draw of a pair $\left(
u_{j},u_{k}\right)  $ gives a distinction of $\pi$, we say the partition $\pi$
\textit{distinguishes}.

The parallelism or duality between logical probabilities and logical entropies
based on the parallel roles of `\textit{its'} (elements of subsets) and
`\textit{dits'} (distinctions of partitions) is summarized in Table 2.

\begin{center}%
\begin{tabular}
[c]{|c||c|c|}\hline
Table 2 & Logical Probability Theory & Logical Information
Theory\\\hline\hline
`Outcomes' & Elements $u\in U$ finite & Dits $\left(  u,u^{\prime}\right)  \in
U\times U$ finite\\\hline
`Events' & Subsets $S\subseteq U$ & Ditsets $\operatorname*{dit}\left(
\pi\right)  \subseteq U\times U$\\\hline
Equiprobable points & $\ \Pr\left(  S\right)  =\frac{|S|}{\left\vert
U\right\vert }$ & $h\left(  \pi\right)  =\frac{\left\vert \operatorname*{dit}%
\left(  \pi\right)  \right\vert }{\left\vert U\times U\right\vert }$\\\hline
Point probabilities & $\ \Pr\left(  S\right)  =\sum\left\{  p_{j}:u_{j}\in
S\right\}  $ & $h\left(  \pi\right)  =\sum\left\{  p_{j}p_{k}:\left(
u_{j},u_{k}\right)  \in\operatorname*{dit}\left(  \pi\right)  \right\}
$\\\hline
Interpretation & $\Pr(S)=$ $1$-draw prob. of $S$-element & $h\left(
\pi\right)  =$ $2$-draw prob. of $\pi$-distinction\\\hline
\end{tabular}

Table 2: Classical logical probability theory and classical logical
information theory
\end{center}

This concludes the argument that logical information theory arises out of
partition logic just as logical probability theory arises out of subset logic.
Now we turn to the formulas of logical information theory and the comparison
to the formulas of Shannon information theory.

\section{History of logical entropy formula}

The formula for logical entropy is not new. Given a finite probability
distribution $p=\left(  p_{1},...,p_{n}\right)  $, the formula $h\left(
p\right)  =1-\sum_{i=1}^{n}p_{i}^{2}$ was used by Gini in 1912
(\cite{gini:vem} reprinted in \cite[p. 369]{gini:vemrpt}) as a measure of
\textquotedblleft mutability\textquotedblright\ or diversity. What is new here
is not the formula, but the derivation from partition logic.

As befits the logical origin of the formula, it occurs in a variety of fields.
The formula in the complementary form, $\sum_{i}p_{i}=1-h\left(  p\right)  $,
was developed early in the $20^{th}$ century in cryptography. The American
cryptologist, William F. Friedman, devoted a 1922 book (\cite{fried:ioc}) to
the index of coincidence (i.e., $\sum p_{i}^{2}$). Solomon Kullback worked as
an assistant to Friedman and wrote a book on cryptology which used the index.
\cite{kull:crypt} During World War II, Alan M. Turing worked for a time in the
Government Code and Cypher School at the Bletchley Park facility in England.
Probably unaware of the earlier work, Turing used $\rho=\sum p_{i}^{2}$ in his
cryptoanalysis work and called it the \textit{repeat rate} since it is the
probability of a repeat in a pair of independent draws from a population with
those probabilities.

After the war, Edward H. Simpson, a British statistician, proposed $\sum
_{B\in\pi}p_{B}^{2}$ as a measure of species concentration (the opposite of
diversity) where $\pi=\left\{  B,B^{\prime},...\right\}  $ is the partition of
animals or plants according to species and where each animal or plant is
considered as equiprobable so $p_{B}=\frac{|B|}{\left\vert U\right\vert }$.
And Simpson gave the interpretation of this homogeneity measure as
\textquotedblleft the probability that two individuals chosen at random and
independently from the population will be found to belong to the same
group.\textquotedblright\cite[p. 688]{simp:md} Hence $1-\sum_{B\in\pi}%
p_{B}^{2}$ is the probability that a random ordered pair will belong to
different species, i.e., will be distinguished by the species partition. In
the biodiversity literature \cite{ric:unify}, the formula $1-\sum_{B\in\pi
}p_{B}^{2}$ is known as \textit{Simpson's index of diversity}\ or sometimes,
the \textit{Gini-Simpson index }\cite{rao:div}.

However, Simpson along with I. J. Good worked at Bletchley Park during WWII,
and, according to Good, \textquotedblleft E. H. Simpson and I both obtained
the notion [the repeat rate] from Turing.\textquotedblright\ \cite[p.
395]{good:turing} When Simpson published the index in 1948, he (again,
according to Good) did not acknowledge Turing \textquotedblleft fearing that
to acknowledge him would be regarded as a breach of
security.\textquotedblright\ \cite[p. 562]{good:div} Since for many purposes
logical entropy offers an alternative to Shannon entropy (\cite{shannon:comm},
\cite{shannonweaver:comm}) in classical information theory, and the quantum
version of logical entropy offers an alternative to von Neumann entropy
\cite{nielsen-chuang:bible} in quantum information theory, it might be useful
to call it `Turing entropy' to have a competitive `famous name' label. But
even before the logical derivation of the formula, I. J. Good pointed out a
certain naturalness:

\begin{quote}
If $p_{1},...,p_{t}$ are the probabilities of $t$ mutually exclusive and
exhaustive events, any statistician of this century who wanted a measure of
homogeneity would have take about two seconds to suggest $\sum p_{i}^{2}$
which I shall call $\rho$.\ \cite[p. 561]{good:div}
\end{quote}

\noindent In view of the frequent and independent discovery and rediscovery of
the formula $\rho=\sum p_{i}^{2}$ or its complement $h(p)=1-\sum p_{i}^{2}$ by
Gini, Friedman, Turing, and many others [e.g., the Hirschman-Herfindahl index
of industrial concentration in economics (\cite{hirsch:pat}, \cite{her:conc}%
)], I. J. Good wisely advises that \textquotedblleft it is unjust to associate
$\rho$ with any one person.\textquotedblright\ \cite[p. 562]{good:div}

\section{Entropy as a \textit{measure} of information}

For a partition $\pi=\left\{  B_{1},...,B_{m}\right\}  $ with block
probabilities $p\left(  B_{i}\right)  $ (obtained using equiprobable points or
with point probabilities), the \textit{Shannon entropy of the partition}
(using natural logs) is:

\begin{center}
$H\left(  \pi\right)  =-\sum_{i=1}^{m}p\left(  B_{i}\right)  \ln\left(
p\left(  B_{i}\right)  \right)  $.
\end{center}

\noindent Or if given a finite probability distribution $p=\left\{
p_{1},...,p_{m}\right\}  $, the \textit{Shannon entropy of the probability
distribution} is:

\begin{center}
$H\left(  p\right)  =-\sum_{i=1}^{m}p_{i}\ln\left(  p_{i}\right)  $.
\end{center}

Shannon entropy and the many other suggested `entropies' are routinely called
\textquotedblleft measures\ of information\textquotedblright%
\ \cite{aczel-d:measures}. The formulas for mutual information, joint entropy,
and conditional entropy are defined so these Shannon entropies satisfy Venn
diagram formulas (\cite[p. 109]{abramson:it}; \cite[p. 508]%
{nielsen-chuang:bible}) that would follow automatically if Shannon entropy
were a measure in the technical sense. As Lorne Campbell put it:

\begin{quotation}
\noindent Certain analogies between entropy and measure have been noted by
various authors. These analogies provide a convenient mnemonic for the various
relations between entropy, conditional entropy, joint entropy, and mutual
information. It is interesting to speculate whether these analogies have a
deeper foundation. It would seem to be quite significant if entropy did admit
an interpretation as the measure of some set. \cite[p. 112]{camp:meas}
\end{quotation}

For any finite set $X$, a \textit{measure} $\mu$ is a function $\mu:\wp\left(
X\right)  \rightarrow%
%TCIMACRO{\U{211d} }%
%BeginExpansion
\mathbb{R}
%EndExpansion
$ such that:

\begin{enumerate}
\item $\mu\left(  \emptyset\right)  =0$,

\item for any $E\subseteq X$, $\mu\left(  E\right)  \geq0$, and

\item for any disjoint subsets $E_{1}$ and $E_{2}$, $\mu(E_{1}\cup E_{2}%
)=\mu\left(  E_{1}\right)  +\mu\left(  E_{2}\right)  $.
\end{enumerate}

Considerable effort has been expended to try to find a framework in which
Shannon entropy would be a measure in this technical sense and thus would
satisfy the desiderata:

\begin{quotation}
\noindent\noindent\noindent that $H\left(  \alpha\right)  $ and $H\left(
\beta\right)  $ are measures of sets, that $H\left(  \alpha,\beta\right)  $ is
the measure of their union, that $I\left(  \alpha,\beta\right)  $ is the
measure of their intersection, and that $H\left(  \alpha|\beta\right)  $ is
the measure of their difference. The possibility that $I\left(  \alpha
,\beta\right)  $ is the entropy of the \textquotedblleft
intersection\textquotedblright\ of two partitions is particularly interesting.
This \textquotedblleft intersection,\textquotedblright\ if it existed, would
presumably contain the information common to the partitions $\alpha$ and
$\beta$.\cite[p. 113]{camp:meas}
\end{quotation}

\noindent But these efforts have not been successful beyond special cases such
as $2^{n}$ equiprobable elements where, as Campbell notes, the Shannon entropy
is just the counting measure $n$ of the minimum number of binary partitions it
takes to distinguish all the elements. In general, Shannon entropy is
\textit{not} a measure.

In contrast, it is \textquotedblleft quite significant\textquotedblright\ that
logical entropy \textit{is} a measure, the normalized counting measure on the
ditset $\operatorname*{dit}(\pi)$ representation of a partition $\pi$ as a
subset of the set $U\times U$. Thus all of Campbell's desiderata are true when:

\begin{itemize}
\item \textquotedblleft sets\textquotedblright\ = ditsets, the set of
distinctions of partitions (or, in general, information sets or infosets), and

\item \textquotedblleft entropies\textquotedblright\ = normalized counting
measure of the ditsets (or, in general, product probability measure on the
infosets), i.e., the logical entropies.
\end{itemize}

The compound Shannon entropy notions satisfy the measure-like formulas, e.g.,
$H(\alpha,\beta)=H\left(  \alpha\right)  +H\left(  \beta\right)  -I\left(
\alpha,\beta\right)  $, not because Shannon entropy is a \textquotedblleft
measure of some set\textquotedblright\ but because logical entropy is such a
measure and all the Shannon compound entropy notions result from the
corresponding logical entropy compound notions by a \textquotedblleft dit-bit
transform\textquotedblright\ that preserves those formulas.

\section{The dit-bit transform}

The logical entropy formulas for various compound notions (e.g., conditional
entropy, mutual information, and joint entropy) stand in certain Venn diagram
relationships \textit{because} logical entropy is a measure. The Shannon
entropy formulas for these compound notions are \textit{defined} so as to
satisfy the Venn diagram relationships \textit{as if} Shannon entropy was a
measure when it is not. How can that be? Perhaps there is some
\textquotedblleft deeper foundation\textquotedblright\ \cite[p. 112]%
{camp:meas} to explain why the Shannon formulas still satisfy those
measure-like Venn diagram relationships.

Indeed, there is such a connection, the dit-bit transform. This transform can
be heuristically motivated by considering two ways to treat the set $U_{n}$ of
$n$ elements with the equal probabilities $p_{0}=\frac{1}{n}$. In that basic
case of an equiprobable set, we can derive the dit-bit connection, and then by
using a probabilistic average, we can develop the Shannon entropy, expressed
in terms of bits, from the logical entropy, expressed in terms of (normalized)
dits, or vice-versa.

Given $U_{n}$ with $n$ equiprobable elements, the number of dits (of the
discrete partition on $U_{n}$) is $n^{2}-n$ so the normalized dit count is:

\begin{center}
$h\left(  p_{0}\right)  =h\left(  \frac{1}{n}\right)  =1-p_{0}=1-\frac{1}{n}$
normalized dits.
\end{center}

\noindent That is the dit-count or logical measure of the information in a set
of $n$ distinct elements (think of it as the logical entropy of the discrete
partition on $U_{n}$ with equiprobable elements).

But we can also measure the information in the set by the number of binary
partitions it takes (on average) to distinguish the elements, and that
bit-count is \cite{hart:ti}:

\begin{center}
$H\left(  p_{0}\right)  =H\left(  \frac{1}{n}\right)  =\log\left(  \frac
{1}{p_{0}}\right)  =\log\left(  n\right)  $ bits.

\textit{Shannon-Hartley entropy for an equiprobable set} $U$ of $n$ elements
\end{center}

The \textit{dit-bit connection} is that the Shannon-Hartley entropy $H\left(
p_{0}\right)  =\log\left(  \frac{1}{p_{0}}\right)  $ will play the same role
in the Shannon formulas that $h\left(  p_{0}\right)  =1-p_{0}$ plays in the
logical entropy formulas--when both are formulated as probabilistic averages.

The common thing being measured is an equiprobable $U_{n}$ where $n=\frac
{1}{p_{0}}$.\footnote{Note that $n=1/p_{0}$ need not be an integer. We are
following the usual practice in information theory where an implicit
\textquotedblleft on average\textquotedblright\ interpretation is assumed
since actual \textquotedblleft binary partitions\textquotedblright\ or
\textquotedblleft binary digits" (or \textquotedblleft bits\textquotedblright)
only come in integral units. The "on average" provisos are justified by the
\textquotedblleft noiseless coding theorem\textquotedblright\ covered in the
later section on the statistical interpretation of Shannon entropy.} The
dit-count for $U_{n}$ is $h\left(  p_{0}\right)  =1-p_{0}$ and the bit-count
for $U$ is $H\left(  p_{0}\right)  =\log\left(  \frac{1}{p_{0}}\right)  $, and
the dit-bit transform converts one count into the other. Using this dit-bit
transform between the two different ways to quantify the `information' in
$U_{n}$, each entropy can be developed from the other. Nevertheless, this
dit-bit connection should not be interpreted as if there is one thing
`information' that can be measured on different scales--like measuring a
length using inches or centimeters. Indeed, the (average) bit-count is a
\textquotedblleft coarser-grid\textquotedblright\ that loses some information
in comparison to the (exact) dit-count as shown by the analysis (below) of
mutual information. There is no bit-count mutual information between
independent probability distributions but there is always dit-count
information even between two (non-trivial) independent distributions (see the
proposition that nonempty ditsets always intersect).

We start with the logical entropy of a probability distribution $p=\left(
p_{1},...,p_{n}\right)  $:

\begin{center}
$h\left(  p\right)  =\sum_{i=1}^{n}p_{i}h\left(  p_{i}\right)  =\sum_{i}%
p_{i}\left(  1-p_{i}\right)  $.
\end{center}

\noindent It is expressed as the probabilistic average of the dit-counts or
logical entropies of the sets $U_{1/p_{i}}$ with $\frac{1}{p_{i}}$
equiprobable elements. But if we switch to the binary-partition bit-counts of
the information content of those same sets $U_{1/p_{i}}$ of $\frac{1}{p_{i}}$
equiprobable elements, then the bit-counts are $H\left(  p_{i}\right)
=\log\left(  \frac{1}{p_{i}}\right)  $ and the probabilistic average is the
Shannon entropy: $H\left(  p\right)  =\sum_{i=1}^{n}p_{i}H\left(
p_{i}\right)  =\sum_{i}p_{i}\log\left(  \frac{1}{p_{i}}\right)  $. Both
entropies have the mathematical form as a probabilistic average or expectation:

\begin{center}
$\sum_{i}p_{i}\left(  \text{amount of `information' in }U_{1/p_{i}}\right)  $
\end{center}

\noindent and differ by using either the dit-count or bit-count conception of
information in $U_{1/p_{i}}$.

The dit-bit connection carries over to all the compound notions of entropy so
that the Shannon notions of conditional entropy, mutual information,
cross-entropy, and divergence can all be developed from the corresponding
notions for logical entropy. Since the logical notions are the values of a
probability measure, the compound notions of logical entropy have the usual
Venn diagram relationships. And then by the dit-bit transform, those Venn
diagram relationships carry over to the compound Shannon formulas since the
dit-bit transform preserves sums and differences (i.e., is, in that sense,
linear). \textit{That} is why the Shannon formulas satisfy the Venn diagram
relationships even though Shannon entropy is not a measure.\footnote{Perhaps,
one should say that Shannon entropy is not the measure of any independently
defined set. The fact that the Shannon formulas `act like a measure' can, of
course, be formalized by formally associating an (indefinite) `set' with each
random variable $X$ and then \textit{defining} the measure value on the `set'
as $H\left(  X\right)  $. Since this `measure' is defined by the Shannon
entropy values, nothing is added to the already-known fact that the Shannon
entropies act like a measure in the Venn diagram relationships. This
formalization seems to have been first carried out by Hu \cite{hu:info} but
was also used by Csiszar and K\"{o}rner \cite{csis-korn:infotheory}, and by
Yeung (\cite{yeung:outlook}; \cite{yeung:firstcourse}).}

Note that while the logical entropy formula $h\left(  p\right)  =\sum_{i}%
p_{i}\left(  1-p_{i}\right)  $ (and the corresponding compound formulas) are
put into that form of an average or expectation to apply the dit-bit
transform, logical entropy is the exact measure of the subset $S_{p}=\left\{
\left(  i,i^{\prime}\right)  :i\neq i^{\prime}\right\}  \subseteq\left\{
1,...,n\right\}  \times\left\{  1,...,n\right\}  $ for the product probability
measure $\mu:\left\{  1,...,n\right\}  ^{2}\rightarrow\left[  0,1\right]  $
where for $S\subseteq\left\{  1,...,n\right\}  ^{2}$, $\mu\left(  S\right)
=\sum\left\{  p_{i}p_{i^{\prime}}:\left(  i,i^{\prime}\right)  \in S\right\}
$, i.e., $h\left(  p\right)  =\mu\left(  S_{p}\right)  $.

\section{Conditional entropies}

\subsection{Logical conditional entropy}

All the compound notions for Shannon and logical entropy could be developed
using either partitions (with point probabilities) or probability
distributions of random variables as the given data. Since the treatment of
Shannon entropy is most often in terms of probability distributions, we will
stick to that case for both types of entropy. The formula for the compound
notion of logical entropy will be developed first, and then the formula for
the corresponding Shannon compound entropy will be obtained by the dit-bit transform.

The general idea of a conditional entropy of a random variable $X$ given a
random variable $Y$ is to measure the additional information in $X$ when we
take away the information contained in $Y$.

Consider a joint probability distribution $\left\{  p\left(  x,y\right)
\right\}  $ on the finite sample space $X\times Y$, with the marginal
distributions $\left\{  p\left(  x\right)  \right\}  $ and $\left\{  p\left(
y\right)  \right\}  $ where $p\left(  x\right)  =\sum_{y\in Y}p\left(
x,y\right)  $ and $p\left(  y\right)  =\sum_{x\in X}p\left(  x,y\right)  $.
For notational simplicity, the entropies can be considered as functions of the
random variables or of their probability distributions, e.g., $h\left(
\left\{  p\left(  x,y\right)  \right\}  \right)  =h\left(  X,Y\right)  $,
$h\left(  \left\{  p\left(  x\right)  \right\}  \right)  =h\left(  X\right)
$, and $h\left(  \left\{  p\left(  y\right)  \right\}  \right)  =h\left(
Y\right)  $. For the joint distribution, we have the:

\begin{center}
$h\left(  X,Y\right)  =\sum_{x\in X,y\in Y}p\left(  x,y\right)  \left[
1-p\left(  x,y\right)  \right]  =1-\sum_{x,y}p\left(  x,y\right)  ^{2}$

\textit{Logical entropy of the joint distribution}
\end{center}

\noindent which is the probability that two samplings of the joint
distribution will yield a pair of \textit{distinct} ordered pairs $\left(
x,y\right)  $, $\left(  x^{\prime},y^{\prime}\right)  \in X\times Y$, i.e.,
with an $X$-distinction $x\neq x^{\prime}$ \textit{or} a $Y$-distinction
$y\neq y^{\prime}$(since ordered pairs are distinct if distinct on one of the
coordinates). The logical entropy notions for the probability distribution
$\left\{  p\left(  x,y\right)  \right\}  $ on $X\times Y$ are all product
probability measures $\mu\left(  S\right)  $ of certain subsets $S\subseteq
\left(  X\times Y\right)  ^{2}$. For the logical entropies defined so far, the
infosets are:

\begin{center}
$S_{X}=\left\{  \left(  \left(  x,y\right)  ,\left(  x^{\prime},y^{\prime
}\right)  \right)  :x\neq x^{\prime}\right\}  $ where $h\left(  X\right)
=\mu\left(  S_{X}\right)  $;

$S_{Y}=\left\{  \left(  \left(  x,y\right)  ,\left(  x^{\prime},y^{\prime
}\right)  \right)  :y\neq y^{\prime}\right\}  $ where $h\left(  Y\right)
=\mu\left(  S_{Y}\right)  $; and

$S_{X\vee Y}=\left\{  \left(  \left(  x,y\right)  ,\left(  x^{\prime
},y^{\prime}\right)  \right)  :x\neq x^{\prime}\vee y\neq y^{\prime}\right\}
=S_{X}\cup S_{Y}$ where $h\left(  X,Y\right)  =\mu\left(  S_{X\vee Y}\right)
=\mu\left(  S_{X}\cup S_{Y}\right)  $.
\end{center}

The infosets $S_{X}$ and $S_{Y}$, as well as their complements $S_{\lnot
X}=\left\{  \left(  \left(  x,y\right)  ,\left(  x^{\prime},y^{\prime}\right)
\right)  :x=x^{\prime}\right\}  $ and $S_{\lnot Y}=\left\{  \left(  \left(
x,y\right)  ,\left(  x^{\prime},y^{\prime}\right)  \right)  :y=y^{\prime
}\right\}  $, generate a Boolean subalgebra $\mathcal{I}\left(  X\times
Y\right)  $ of $\wp\left(  \left(  X\times Y\right)  \times\left(  X\times
Y\right)  \right)  $ which might be called the \textit{information algebra of
}$X\times Y$. It is defined independently of any probability measure $\left\{
p\left(  x,y\right)  \right\}  $ on $X\times Y$, and any such measure defines
the product measure $\mu$ on $\left(  X\times Y\right)  \times\left(  X\times
Y\right)  $, and the corresponding logical entropies are the product measures
on the infosets in $\mathcal{I}\left(  X\times Y\right)  $.

For the definition of the conditional entropy $h\left(  X|Y\right)  $, we
simply take the product measure of the set of pairs $\left(  x,y\right)  $ and
$\left(  x^{\prime},y^{\prime}\right)  $ that give an $X$-distinction but not
a $Y$-distinction. Hence we use the inequation $x\neq x^{\prime}$ for the
$X$-distinction and negate the $Y$-distinction $y\neq y^{\prime}$ to get the
infoset that is the difference of the infosets for $X$ and $Y$:

\begin{center}
$S_{X\wedge\lnot Y}=\left\{  \left(  \left(  x,y\right)  ,\left(  x^{\prime
},y^{\prime}\right)  \right)  :x\neq x^{\prime}\wedge y=y^{\prime}\right\}
=S_{X}-S_{Y}$ so 

$h\left(  X|Y\right)  =\mu\left(  S_{X\wedge\lnot Y}\right)  =\mu\left(
S_{X}-S_{Y}\right)  $.
\end{center}

\noindent Since $S_{X\vee Y}=S_{X\wedge\lnot Y}\uplus S_{Y}$ and the union is
disjoint, we have for the measure $\mu$:

\begin{center}
$h\left(  X,Y\right)  =\mu\left(  S_{X\vee Y}\right)  =\mu\left(
S_{X\wedge\lnot Y}\right)  +\mu\left(  S_{Y}\right)  =h\left(  X|Y\right)
+h\left(  Y\right)  $,
\end{center}

\noindent which is illustrated in the Venn diagram Figure 1.%

%TCIMACRO{\FRAME{dtbpF}{2.3886in}{1.8135in}{0pt}{}{}{figure1.eps}%
%width 2.3886in;  height 1.8135in;  depth 0pt;  original-width 3.2206in;
%original-height 2.4388in;  cropleft "0";  croptop "1";  cropright "1";
%cropbottom "0";  filename 'figure1.eps';file-properties "XNPEU";}} }%
%BeginExpansion
\begin{center}
\includegraphics[
height=1.8135in,
width=2.3886in
]%
{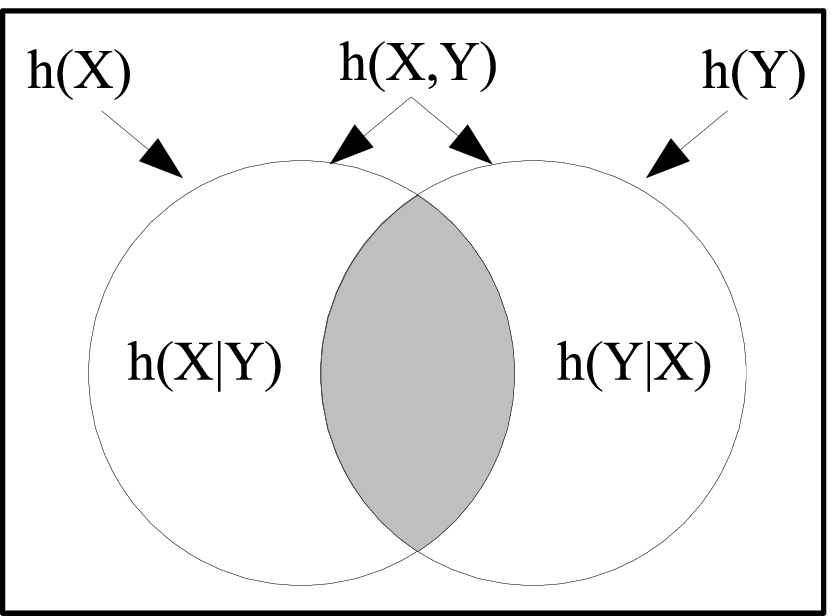}%
\end{center}
%EndExpansion

\begin{center}
Figure 1: $h\left(  X,Y\right)  =h\left(  X|Y\right)  +h\left(  Y\right)  $.
\end{center}

\noindent In terms of the probabilities:

\begin{center}
$h\left(  X|Y\right)  =h\left(  X,Y\right)  -h\left(  Y\right)  =\sum
_{x,y}p\left(  x,y\right)  \left(  1-p\left(  x,y\right)  \right)  -\sum
_{y}p\left(  y\right)  \left(  1-p\left(  y\right)  \right)  $

$\sum_{x,y}p\left(  x,y\right)  \left[  \left(  1-p\left(  x,y\right)
\right)  -\left(  1-p\left(  y\right)  \right)  \right]  $

\textit{Logical conditional entropy of }$X$\textit{\ given }$Y$.
\end{center}

\subsection{Shannon conditional entropy}

Given the joint distribution $\left\{  p\left(  x,y\right)  \right\}  $ on
$X\times Y$, the conditional probability distribution for a specific $y_{0}\in
Y$ is $p\left(  x|y_{0}\right)  =\frac{p\left(  x,y_{0}\right)  }{p\left(
y_{0}\right)  }$ which has the Shannon entropy: $H\left(  X|y_{0}\right)
=\sum_{x}p\left(  x|y_{0}\right)  \log\left(  \frac{1}{p\left(  x|y_{0}%
\right)  }\right)  $. Then the Shannon conditional entropy is defined as the
\textit{average} of these entropies:

\begin{center}
$H\left(  X|Y\right)  =\sum_{y}p\left(  y\right)  \sum_{x}\frac{p\left(
x,y\right)  }{p\left(  y\right)  }\log\left(  \frac{p\left(  y\right)
}{p\left(  x,y\right)  }\right)  =\sum_{x,y}p\left(  x,y\right)  \log\left(
\frac{p\left(  y\right)  }{p\left(  x,y\right)  }\right)  $

\textit{Shannon conditional entropy of }$X$\textit{\ given }$Y$.
\end{center}

\noindent All the Shannon notions can be obtained by the dit-bit transform of
the corresponding logical notions. Applying the transform $1-p\rightsquigarrow
\log\left(  \frac{1}{p}\right)  $ to the logical conditional entropy expressed
as an average of \textquotedblleft$1-p$\textquotedblright\ expressions:
$h\left(  X|Y\right)  =\sum_{x,y}p\left(  x,y\right)  \left[  \left(
1-p\left(  x,y\right)  \right)  -\left(  1-p\left(  y\right)  \right)
\right]  $, yields the Shannon conditional entropy:

\begin{center}
$H\left(  X|Y\right)  =\sum_{x,y}p\left(  x,y\right)  \left[  \log\left(
\frac{1}{p\left(  x,y\right)  }\right)  -\log\left(  \frac{1}{p\left(
y\right)  }\right)  \right]  =\sum_{x,y}p\left(  x,y\right)  \log\left(
\frac{p\left(  y\right)  }{p\left(  x,y\right)  }\right)  $.
\end{center}

\noindent Since the dit-bit transform preserves sums and differences, we will
have the same sort of Venn diagram formula for the Shannon entropies (even
though the Shannon notions are not the values of a measure) and this can be
illustrated in a similar \textquotedblleft mnemonic\textquotedblright\ Venn diagram.%

%TCIMACRO{\FRAME{dtbpF}{2.3289in}{1.8464in}{0pt}{}{}{figure2.eps}%
%width 2.3289in;  height 1.8464in;  depth 0pt;  original-width 3.0822in;
%original-height 2.4388in;  cropleft "0";  croptop "1";  cropright "1";
%cropbottom "0";  filename 'figure2.eps';file-properties "XNPEU";}} }%
%BeginExpansion
\begin{center}
\includegraphics[
height=1.8464in,
width=2.3289in
]%
{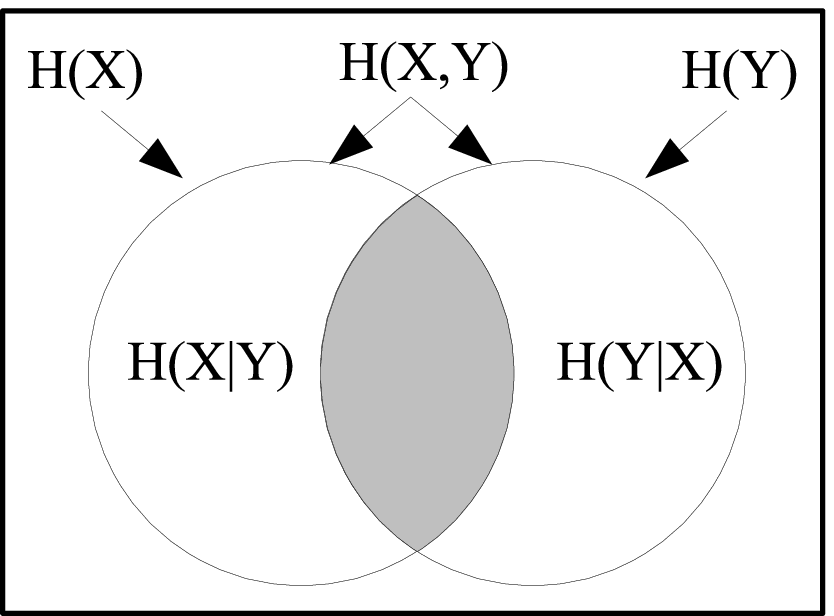}%
\end{center}
%EndExpansion

\begin{center}
Figure 2: $H\left(  X|Y\right)  =H\left(  X,Y\right)  -H\left(  Y\right)  $.
\end{center}

\section{Mutual information}

\subsection{Logical mutual information}

Intuitively, the mutual logical information $m\left(  X,Y\right)  $ in the
joint distribution$\left\{  p\left(  x,y\right)  \right\}  $ would be the
probability that a sampled pair of pairs $\left(  x,y\right)  $ and $\left(
x^{\prime},y^{\prime}\right)  $ would be distinguished in \textit{both}
coordinates, i.e., a distinction $x\neq x^{\prime}$ of $p\left(  x\right)  $
\textit{and }a distinction $y\neq y^{\prime}$ of $p\left(  y\right)  $. In
terms of subsets, the subset for the mutual information is intersection of
infosets for $X$ and $Y$:

\begin{center}
$S_{X\wedge Y}=S_{X}\cap S_{Y}$ so $m\left(  X,Y\right)  =\mu\left(
S_{X\wedge Y}\right)  =\mu\left(  S_{X}\cap S_{Y}\right)  $.
\end{center}

In terms of disjoint unions of subsets:

\begin{center}
$S_{X\vee Y}=S_{X\wedge\lnot Y}\uplus S_{Y\wedge\lnot X}\uplus S_{X\wedge Y}$
\end{center}

\noindent so

\begin{center}
$h\left(  X,Y\right)  =\mu\left(  S_{X\vee Y}\right)  =\mu\left(
S_{X\wedge\lnot Y}\right)  +\mu\left(  S_{Y\wedge\lnot X}\right)  +\mu\left(
S_{X\wedge Y}\right)  $

$=h\left(  X|Y\right)  +h\left(  Y|X\right)  +m\left(  X,Y\right)  $ (as in
Figure 3),
\end{center}

\noindent or:

\begin{center}
$m\left(  X,Y\right)  =h\left(  X\right)  +h\left(  Y\right)  -h\left(
X,Y\right)  $.
\end{center}

%

%TCIMACRO{\FRAME{dtbpF}{2.2321in}{1.6423in}{0pt}{}{}{figure3.eps}%
%width 2.2321in;  height 1.6423in;  depth 0pt;  original-width 3.3053in;
%original-height 2.4258in;  cropleft "0";  croptop "1";  cropright "1";
%cropbottom "0";  filename 'figure3.eps';file-properties "XNPEU";}} }%
%BeginExpansion
\begin{center}
\includegraphics[
height=1.6423in,
width=2.2321in
]%
{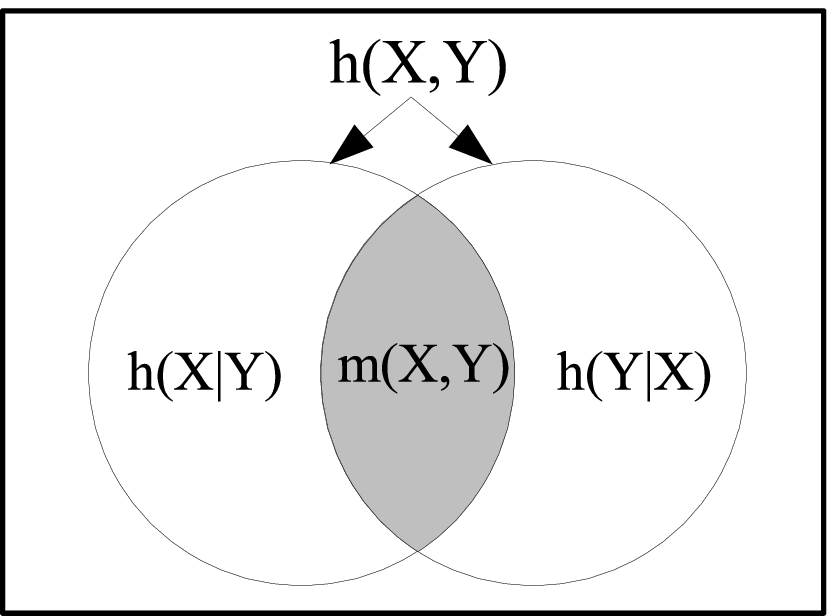}%
\end{center}
%EndExpansion

\begin{center}
Figure 3: $h\left(  X,Y\right)  =h\left(  X|Y\right)  +h\left(  Y|X\right)
+m\left(  X,Y\right)  $
\end{center}

Expanding $m\left(  X,Y\right)  =h\left(  X\right)  +h\left(  Y\right)
-h\left(  X,Y\right)  $ in terms of probability averages gives:

\begin{center}
$m\left(  X,Y\right)  =\sum_{x,y}p\left(  x,y\right)  \left[  \left[
1-p\left(  x\right)  \right]  +\left[  1-p\left(  y\right)  \right]  -\left[
1-p\left(  x,y\right)  \right]  \right]  $

\textit{Logical mutual information in a joint probability distribution}.
\end{center}

It is a non-trivial fact that nonempty ditsets \textit{always} intersect. It
helps to sharpen the differences between logical and Shannon entropies in the
matter of `intuitions' about independence (see below). To prove the result for
any joint probability distributions $\left\{  p\left(  x,y\right)  \right\}  $
on the finite set $X\times Y$, the corresponding `ditsets' for $X$ and $Y$ are
the possible supports for the infosets $S_{X}$ and $S_{Y}$:

\begin{center}
$\operatorname*{dit}\left(  X\right)  =\left\{  \left(  \left(  x,y\right)
,\left(  x^{\prime},y^{\prime}\right)  \right)  :x\neq x^{\prime},p\left(
x,y\right)  p\left(  x^{\prime},y^{\prime}\right)  >0\right\}  \subseteq
\left(  X\times Y\right)  ^{2}$

$\operatorname*{dit}\left(  Y\right)  =\left\{  \left(  \left(  x,y\right)
,\left(  x^{\prime},y^{\prime}\right)  \right)  :y\neq y^{\prime},p\left(
x,y\right)  p\left(  x^{\prime},y^{\prime}\right)  >0\right\}  \subseteq
\left(  X\times Y\right)  ^{2}$.
\end{center}

Now $\operatorname*{dit}\left(  X\right)  \subseteq S_{X}$ and
$\operatorname*{dit}\left(  Y\right)  \subseteq S_{Y}$, and for the product
probability measure $\mu$ on $\left(  X\times Y\right)  ^{2}$, the sets
$S_{X}-\operatorname*{dit}\left(  X\right)  $ and $S_{Y}-\operatorname*{dit}%
\left(  Y\right)  $ are of measure $0$ so:

\begin{center}
$\mu\left(  \operatorname*{dit}\left(  X\right)  \right)  =\mu\left(
S_{X}\right)  =h\left(  X\right)  $
\end{center}

\noindent and similarly $\mu\left(  \operatorname*{dit}\left(  Y\right)
\right)  =h\left(  Y\right)  $. Then $h\left(  X\right)  =0$ iff
$\operatorname*{dit}\left(  X\right)  =\emptyset$ iff there is an $x_{0}\in X$
such that $p\left(  x_{0}\right)  =1$, and similarly for $h\left(  Y\right)  $.

\begin{proposition}
[Nonempty ditsets always intersect]If $h\left(  X\right)  h\left(  Y\right)
>0$, then $m\left(  X,Y\right)  >0$.
\end{proposition}

Proof: Since $\operatorname*{dit}\left(  X\right)  $ is nonempty, there are
two pairs $\left(  x,y\right)  $ and $\left(  x^{\prime},y^{\prime}\right)  $
such that $x\neq x^{\prime}$ and $p\left(  x,y\right)  p\left(  x^{\prime
},y^{\prime}\right)  >0$. If $y\neq y^{\prime}$ then $\left(  \left(
x,y\right)  ,\left(  x^{\prime},y^{\prime}\right)  \right)  \in
\operatorname*{dit}\left(  Y\right)  $ as well and we are finished, i.e.,
$\operatorname*{dit}\left(  X\right)  \cap\operatorname*{dit}\left(  Y\right)
\neq\emptyset$. Hence assume $y=y^{\prime}$. Since $\operatorname*{dit}\left(
Y\right)  $ is also nonempty and thus $p\left(  y\right)  \neq1$, there is
another $y^{\prime\prime}$ such that for some $x^{\prime\prime}$, $p\left(
x^{\prime\prime},y^{\prime\prime}\right)  >0$. Since $x^{\prime\prime}$ can't
be equal to both $x$ and $x^{\prime}$, at least one of the pairs $\left(
\left(  x,y\right)  ,\left(  x^{\prime\prime},y^{\prime\prime}\right)
\right)  $ or $\left(  \left(  x^{\prime},y\right)  ,\left(  x^{\prime\prime
},y^{\prime\prime}\right)  \right)  $ is in both $\operatorname*{dit}\left(
X\right)  $ and $\operatorname*{dit}\left(  Y\right)  $, and thus the product
measure on $S_{\wedge\left\{  X,Y\right\}  }=\left\{  \left(  \left(
x,y\right)  ,\left(  x^{\prime},y^{\prime}\right)  \right)  :x\neq x^{\prime
}\wedge y\neq y^{\prime}\right\}  $ is positive, i.e., $m\left(  X,Y\right)
>0$.$\square$

\begin{corollary}
Nonempty infosets $S_{X}$ and $S_{Y}$ always intersect.
\end{corollary}

Proof: For the uniform distribution on $X\times Y$, $\operatorname*{dit}%
\left(  X\right)  =S_{X}$ and $\operatorname*{dit}\left(  Y\right)  =S_{Y}%
$.$\square$

Note that compound infosets like $S_{X\wedge\lnot Y}$ and $S_{X\wedge Y}$ do
not intersect.

\subsection{Shannon mutual information}

Applying the dit-bit transform $1-p\rightsquigarrow\log\left(  \frac{1}%
{p}\right)  $ to the logical mutual information formula

\begin{center}
$m\left(  X,Y\right)  =\sum_{x,y}p\left(  x,y\right)  \left[  \left[
1-p\left(  x\right)  \right]  +\left[  1-p\left(  y\right)  \right]  -\left[
1-p\left(  x,y\right)  \right]  \right]  $
\end{center}

\noindent expressed in terms of probability averages gives the corresponding
Shannon notion:

\begin{center}
$I\left(  X,Y\right)  =\sum_{x,y}p\left(  x,y\right)  \left[  \left[
\log\left(  \frac{1}{p\left(  x\right)  }\right)  \right]  +\left[
\log\left(  \frac{1}{p\left(  y\right)  }\right)  \right]  -\left[
\log\left(  \frac{1}{p\left(  x,y\right)  }\right)  \right]  \right]  $

$=\sum_{x,y}p\left(  x,y\right)  \log\left(  \frac{p\left(  x,y\right)
}{p\left(  x\right)  p\left(  y\right)  }\right)  $

\textit{Shannon mutual information in a joint probability distribution}.
\end{center}

Since the dit-bit transform preserves sums and differences, the logical
formulas for the measures gives the mnemonic Figure 4:

\begin{center}
$I\left(  X,Y\right)  =H\left(  X\right)  +H\left(  Y\right)  -H\left(
X,Y\right)  =H\left(  X,Y\right)  -H\left(  X|Y\right)  -H\left(  Y|X\right)
$.%

%TCIMACRO{\FRAME{dtbpF}{2.4491in}{1.8023in}{0pt}{}{}{figure4.eps}%
%width 2.4491in;  height 1.8023in;  depth 0pt;  original-width 3.3053in;
%original-height 2.4258in;  cropleft "0";  croptop "1";  cropright "1";
%cropbottom "0";  filename 'figure4.eps';file-properties "XNPEU";}} }%
%BeginExpansion
\begin{center}
\includegraphics[
height=1.8023in,
width=2.4491in
]%
{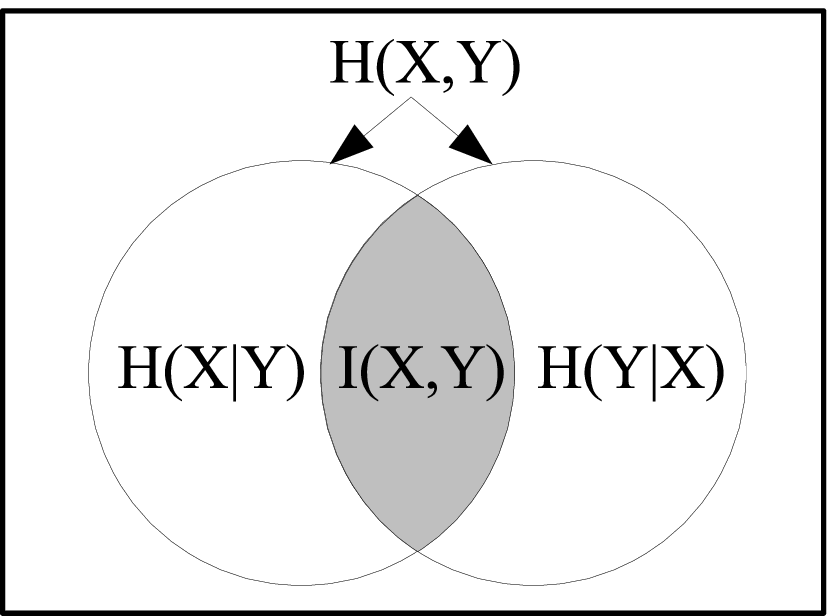}%
\end{center}
%EndExpansion

Figure 4: $H\left(  X,Y\right)  =H\left(  X|Y\right)  +H\left(  Y|X\right)
+I\left(  X,Y\right)  $.
\end{center}

This is the usual Venn diagram for the Shannon entropy notions that needs to
be explained--since the Shannon entropies are not measures. Of course, one
could just say the relationship holds for the Shannon entropies because that's
how they were defined. It may seem a happy accident that the Shannon
definitions all satisfy the measure-like Venn diagram formulas, but as one
author put it: \textquotedblleft Shannon carefully contrived for this
`accident' to occur\textquotedblright\ \cite[p. 153]{roseboom:abstract}. As
noted above, Campbell asked if \textquotedblleft these analogies have a deeper
foundation\textquotedblright\ \cite[p. 112]{camp:meas} and the dit-bit
transform answers that question.

\section{Independent Joint Distributions}

A joint probability distribution $\left\{  p\left(  x,y\right)  \right\}  $ on
$X\times Y$ is \textit{independent} if each value is the product of the
marginals: $p\left(  x,y\right)  =p\left(  x\right)  p\left(  y\right)  $.

For an independent distribution, the Shannon mutual information

\begin{center}
$I\left(  X,Y\right)  =\sum_{x\in X,y\in Y}p\left(  x,y\right)  \log\left(
\frac{p\left(  x,y\right)  }{p\left(  x\right)  p\left(  y\right)  }\right)  $
\end{center}

\noindent is immediately seen to be zero so we have:

\begin{center}
$H\left(  X,Y\right)  =H\left(  X\right)  +H\left(  Y\right)  $

Shannon entropies for independent $\left\{  p\left(  x,y\right)  \right\}  $.
\end{center}

For the logical mutual information $m(X,Y)$, independence gives:%

\begin{align*}
m\left(  X,Y\right)   &  =%
%TCIMACRO{\tsum \nolimits_{x,y}}%
%BeginExpansion
{\textstyle\sum\nolimits_{x,y}}
%EndExpansion
p\left(  x,y\right)  \left[  1-p\left(  x\right)  -p\left(  y\right)
+p\left(  x,y\right)  \right] \\
&  =%
%TCIMACRO{\tsum \nolimits_{x,y}}%
%BeginExpansion
{\textstyle\sum\nolimits_{x,y}}
%EndExpansion
p\left(  x\right)  p\left(  y\right)  \left[  1-p\left(  x\right)  -p\left(
y\right)  +p\left(  x\right)  p\left(  y\right)  \right] \\
&  =%
%TCIMACRO{\tsum \nolimits_{x}}%
%BeginExpansion
{\textstyle\sum\nolimits_{x}}
%EndExpansion
p\left(  x\right)  \left[  1-p\left(  x\right)  \right]
%TCIMACRO{\tsum \nolimits_{y}}%
%BeginExpansion
{\textstyle\sum\nolimits_{y}}
%EndExpansion
p\left(  y\right)  \left[  1-p\left(  y\right)  \right] \\
&  =h\left(  X\right)  h\left(  Y\right)
\end{align*}

\begin{center}
Logical entropies for independent $\left\{  p\left(  x,y\right)  \right\}  $.
\end{center}

The logical conditional entropy $h\left(  X|Y\right)  =h\left(  X,Y\right)
-h\left(  Y\right)  =h\left(  X\right)  -m\left(  X,Y\right)  $ is the
probability that a random pair of pairs $\left(  x,y\right)  $ and $\left(
x^{\prime},y^{\prime}\right)  $ is a distinction $x\neq x^{\prime}$ for
$\left\{  p\left(  x\right)  \right\}  $ but not a distinction $y\neq
y^{\prime}$ of $\left\{  p\left(  y\right)  \right\}  $. Under independence,
that logical conditional entropy is $h\left(  X|Y\right)  =h\left(  X\right)
\left(  1-h\left(  Y\right)  \right)  $ which is the probability of randomly
drawing a distinction from the marginal distribution $\left\{  p\left(
x\right)  \right\}  $ times the probability of randomly drawing an
indistinction from the other marginal distribution $\left\{  p\left(
y\right)  \right\}  $.

The nonempty-ditsets-always-intersect proposition shows that $h\left(
X\right)  h\left(  Y\right)  >0$ implies $m\left(  X,Y\right)  >0$, and thus
that logical mutual information $m\left(  X,Y\right)  $ is still positive for
independent distributions when $h\left(  X\right)  h\left(  Y\right)  >0$, in
which case $m\left(  X,Y\right)  =h\left(  X\right)  h\left(  Y\right)  $.
This is a striking difference between the average bit-count Shannon entropy
and the dit-count logical entropy. Aside from the waste case where $h\left(
X\right)  h\left(  Y\right)  =0$, there are always positive probability mutual
distinctions for $X$ and $Y$, and that dit-count information is not recognized
by the average bit-count Shannon entropy.

\section{Cross-entropies and divergences}

Given two probability distributions $p=\left(  p_{1},...,p_{n}\right)  $ and
$q=\left(  q_{1},...,q_{n}\right)  $ on the same sample space $\left\{
1,...,n\right\}  $, we can again consider the drawing of a pair of points but
where the first drawing is according to $p$ and the second drawing according
to $q$. The probability that the points are distinct would be a natural and
more general notion of logical entropy that would be the:

\begin{center}
$h\left(  p\Vert q\right)  =\sum_{i}p_{i}(1-q_{i})=1-\sum_{i}p_{i}q_{i}$

\textit{Logical} \textit{cross entropy of }$p$ \textit{and} $q$
\end{center}

\noindent which is symmetric. The logical cross entropy is the same as the
logical entropy when the distributions are the same, i.e., if $p=q$, then
$h\left(  p\Vert q\right)  =h\left(  p\right)  $.

Although the logical cross entropy formula is symmetrical in $p$ and $q$,
there are two different ways to express it as an average in order to apply the
dit-bit transform: $\sum_{i}p_{i}(1-q_{i})$ and $\sum_{i}q_{i}\left(
1-p_{i}\right)  $. The two transforms are the two asymmetrical versions of
Shannon cross entropy:

\begin{center}
$H\left(  p\Vert q\right)  =\sum_{i}p_{i}\log\left(  \frac{1}{q_{i}}\right)  $
and $H\left(  q||p\right)  =\sum_{i}q_{i}\log\left(  \frac{1}{p_{i}}\right)  $.
\end{center}

\noindent which is not symmetrical due to the asymmetric role of the
logarithm, although if $p=q$, then $H\left(  p\Vert q\right)  =H\left(
p\right)  $. When the logical cross entropy is expressed as an average in a
symmetrical way: $h\left(  p||q\right)  =\frac{1}{2}\left[  \sum_{i}%
p_{i}(1-q_{i})+\sum_{i}q_{i}\left(  1-p_{i}\right)  \right]  $, then the
dit-bit transform is the \textit{symmetrized Shannon cross entropy}:

\begin{center}
$H_{s}\left(  p||q\right)  =\frac{1}{2}\left[  H\left(  p||q\right)  +H\left(
q||p\right)  \right]  $.
\end{center}

The \textit{Kullback-Leibler divergence }(or \textit{relative entropy})
$D\left(  p\Vert q\right)  =\sum_{i}p_{i}\log\left(  \frac{p_{i}}{q_{i}%
}\right)  $ is defined as a measure of the distance or divergence between the
two distributions where $D\left(  p\Vert q\right)  =H\left(  p\Vert q\right)
-H\left(  p\right)  $. A basic result is the:

\begin{center}
$D\left(  p\Vert q\right)  \geq0$ with equality if and only if $p=q$

\textit{Information inequality} \cite[p. 26]{cover:eit}.
\end{center}

The \textit{symmetrized Kullback-Leibler divergence} is:

\begin{center}
$D_{s}(p||q)=\frac{1}{2}\left[  D\left(  p||q\right)  +D\left(  q||p\right)
\right]  =H_{s}\left(  p||q\right)  -\left[  \frac{H\left(  p\right)
+H\left(  q\right)  }{2}\right]  $.
\end{center}

But starting afresh, one might ask: \textquotedblleft What is the natural
measure of the difference or distance between two probability distributions
$p=\left(  p_{1},...,p_{n}\right)  $ and $q=\left(  q_{1},...,q_{n}\right)  $
that would always be non-negative, and would be zero if and only if they are
equal?\textquotedblright\ The (Euclidean) distance between the two points in $%
%TCIMACRO{\U{211d} }%
%BeginExpansion
\mathbb{R}
%EndExpansion
^{n}$ would seem to be the logical\ answer---so we take that distance (squared
with a scale factor) as the definition of the:

\begin{center}
$d\left(  p\Vert q\right)  =$ $\frac{1}{2}\sum_{i}\left(  p_{i}-q_{i}\right)
^{2}$

\textit{Logical divergence} (or \textit{logical} \textit{relative
entropy})\footnote{In \cite{ell:countingdits}, this definition was given
without the useful scale factor of $1/2$.}
\end{center}

\noindent which is symmetric and we trivially have:

\begin{center}
$d\left(  p||q\right)  \geq0$ with equality iff $p=q$

\textit{Logical information inequality}.
\end{center}

We have component-wise:

\begin{center}
$0\leq\left(  p_{i}-q_{i}\right)  ^{2}=p_{i}^{2}-2p_{i}q_{i}+q_{i}%
^{2}=2\left[  \frac{1}{n}-p_{i}q_{i}\right]  -\left[  \frac{1}{n}-p_{i}%
^{2}\right]  -\left[  \frac{1}{n}-q_{i}^{2}\right]  $
\end{center}

\noindent so that taking the sum for $i=1,...,n$ gives:

\begin{center}%
\begin{align*}
d\left(  p\Vert q\right)   &  =\frac{1}{2}%
%TCIMACRO{\tsum \nolimits_{i}}%
%BeginExpansion
{\textstyle\sum\nolimits_{i}}
%EndExpansion
\left(  p_{i}-q_{i}\right)  ^{2}\\
&  =\left[  1-%
%TCIMACRO{\tsum \nolimits_{i}}%
%BeginExpansion
{\textstyle\sum\nolimits_{i}}
%EndExpansion
p_{i}q_{i}\right]  -\frac{1}{2}\left[  \left(  1-%
%TCIMACRO{\tsum \nolimits_{i}}%
%BeginExpansion
{\textstyle\sum\nolimits_{i}}
%EndExpansion
p_{i}^{2}\right)  +\left(  1-%
%TCIMACRO{\tsum \nolimits_{i}}%
%BeginExpansion
{\textstyle\sum\nolimits_{i}}
%EndExpansion
q_{i}^{2}\right)  \right] \\
&  =h\left(  p\Vert q\right)  -\frac{h\left(  p\right)  +h\left(  q\right)
}{2}\text{.}%
\end{align*}

Logical divergence = \textit{Jensen difference} \cite[p. 25]{rao:div} between
probability distributions.
\end{center}

\noindent Then the information inequality implies that the logical
cross-entropy is greater than or equal to the average of the logical entropies:

\begin{center}
$h\left(  p||q\right)  \geq\frac{h\left(  p\right)  +h\left(  q\right)  }{2}$
with equality iff $p=q$.
\end{center}

The half-and-half probability distribution $\frac{p+q}{2}$ that mixes $p$ and
$q$ has the logical entropy of

\begin{center}
$h\left(  \frac{p+q}{2}\right)  =\frac{h\left(  p\Vert q\right)  }{2}%
+\frac{h\left(  p\right)  +h\left(  q\right)  }{4}=\frac{1}{2}\left[  h\left(
p||q\right)  +\frac{h\left(  p\right)  +h\left(  q\right)  }{2}\right]  $
\end{center}

\noindent so that:

\begin{center}
$h(p||q)\geq h\left(  \frac{p+q}{2}\right)  \geq\frac{h\left(  p\right)
+h\left(  q\right)  }{2}$ with equality iff $p=q$.

Mixing different $p$ and $q$ increases logical entropy.
\end{center}

The logical divergence can be expressed in the proper symmetrical form of
averages to apply the dit-bit transform:

\begin{center}
$d\left(  p\Vert q\right)  =\frac{1}{2}\left[  \sum_{i}p_{i}\left(
1-q_{i}\right)  +\sum_{i}q_{i}\left(  1-p_{i}\right)  \right]  -\frac{1}%
{2}\left[  \left(  \sum_{i}p_{i}\left(  1-p_{i}\right)  \right)  +\left(
\sum_{i}q_{i}\left(  1-q_{i}\right)  \right)  \right]  $
\end{center}

\noindent so the transform is:

\begin{center}
$\frac{1}{2}\left[  \sum_{i}p_{i}\log\left(  \frac{1}{q_{i}}\right)  +\sum
_{i}q_{i}\log\left(  \frac{1}{p_{i}}\right)  -\sum_{i}p_{i}\log\left(
\frac{1}{p_{i}}\right)  -\sum_{i}q_{i}\log\left(  \frac{1}{q_{i}}\right)
\right]  $

$=\frac{1}{2}\left[  \sum_{i}p_{i}\log\left(  \frac{p_{i}}{q_{i}}\right)
+\sum_{i}q_{i}\log\left(  \frac{q_{i}}{p_{i}}\right)  \right]  =\frac{1}%
{2}\left[  D\left(  p||q\right)  +D\left(  q||p\right)  \right]  $

$=D_{s}\left(  p||q\right)  $.
\end{center}

\noindent Since the logical divergence $d\left(  p||q\right)  $ is
symmetrical, it develops via the dit-bit transform to the \textit{symmetrized}
version $D_{s}\left(  p||q\right)  $ of the Kullback-Leibler divergence.

\section{Summary of formulas and dit-bit transforms}

The following table 3 summarizes the concepts for the Shannon and logical
entropies. We use the abbreviations $p_{xy}=p(x,y)$, $p_{x}=p(x)$, and
$p_{y}=p\left(  y\right)  $.

\begin{center}%
\begin{tabular}
[c]{l|c|c|}\cline{2-3}%
Table 3 & $\text{Shannon Entropy}$ & $\text{Logical Entropy}$\\\hline\hline
\multicolumn{1}{|l|}{{\small Entropy}} & $H(p)=\sum p_{i}\log\left(
1/p_{i}\right)  $ & $h\left(  p\right)  =\sum p_{i}\left(  1-p_{i}\right)
$\\\hline
\multicolumn{1}{|l|}{{\small Mutual Info.}} & ${\small I(X,Y)=H}\left(
X\right)  +H\left(  Y\right)  -H\left(  X,Y\right)  $ & $m\left(  X,Y\right)
=h\left(  X\right)  +h\left(  Y\right)  -h\left(  X,Y\right)  $\\\hline
\multicolumn{1}{|l|}{{\small Cond. entropy}} & \ $H\left(  X|Y\right)
=H(X)-I\left(  X,Y\right)  $ & $h\left(  X|Y\right)  =h\left(  X\right)
-m\left(  X,Y\right)  $\\\hline
\multicolumn{1}{|l|}{{\small Independence}} & $I\left(  X,Y\right)  =0$ &
$m\left(  X,Y\right)  =h\left(  X\right)  h\left(  Y\right)  $\\\hline
\multicolumn{1}{|l|}{{\small Indep. Relations}} & $H\left(  X|Y\right)
=H\left(  X\right)  $ & $h\left(  X|Y\right)  =h\left(  X\right)  \left(
1-h\left(  Y\right)  \right)  ${\small \ }\\\hline
\multicolumn{1}{|l|}{{\small Cross entropy}} & ${\small H}\left(  p\Vert
q\right)  =\sum p_{i}\log\left(  1/q_{i}\right)  $ & ${\small h}\left(  p\Vert
q\right)  =\sum p_{i}\left(  1-q_{i}\right)  $\\\hline
\multicolumn{1}{|l|}{{\small Divergence}} & ${\small D}\left(  p\Vert
q\right)  =\sum_{i}p_{i}\log\left(  \frac{p_{i}}{q_{i}}\right)  $ &
{\small \ }${\small d}\left(  p||q\right)  =\frac{1}{2}\sum_{i}\left(
p_{i}-q_{i}\right)  ^{2}$\\\hline
\multicolumn{1}{|l|}{{\small Relationships}} & ${\small D}\left(  p\Vert
q\right)  ={\small H}\left(  p\Vert q\right)  {\small -H}\left(  p\right)  $ &
${\small d}\left(  p\Vert q\right)  ={\small h}\left(  p\Vert q\right)
{\small -}\left[  {\small h}\left(  p\right)  {\small +h}\left(  q\right)
\right]  /2$\\\hline
\multicolumn{1}{|l|}{{\small Info. Inequality}} & ${\small D}\left(  p\Vert
q\right)  \geq{\small 0}\text{ with }=\text{ iff }p=q$ & $d\left(  p\Vert
q\right)  \geq0\text{ with }=\text{ iff }p=q$\\\hline
\end{tabular}

Table 3: Comparisons between Shannon and logical entropy formulas
\end{center}

The following table 4 summarizes the dit-bit transforms.

\begin{center}%
\begin{tabular}
[c]{c|c|}\cline{2-2}%
Table 4 & The Dit-Bit Transform: $1-p_{i}\rightarrow\log\left(  \frac{1}%
{p_{i}}\right)  $\\\hline\hline
\multicolumn{1}{|r|}{$h\left(  p\right)  =$} & $\sum_{i}p_{i}\left(
1-p_{i}\right)  $\\\hline
\multicolumn{1}{|r|}{$H\left(  p\right)  =$} & $\sum_{i}p_{i}\log\left(
1/p_{i}\right)  $\\\hline\hline
\multicolumn{1}{|r|}{$h\left(  X|Y\right)  =$} & $\sum_{x,y}p\left(
x,y\right)  \left[  \left(  1-p\left(  x,y\right)  \right)  -\left(
1-p\left(  y\right)  \right)  \right]  $\\\hline
\multicolumn{1}{|r|}{$H\left(  X|Y\right)  =$} & $\sum_{x,y}p\left(
x,y\right)  \left[  \log\left(  \frac{1}{p\left(  x,y\right)  }\right)
-\log\left(  \frac{1}{p\left(  y\right)  }\right)  \right]  $\\\hline\hline
\multicolumn{1}{|r|}{$m\left(  X,Y\right)  =$} & $\sum_{x,y}p\left(
x,y\right)  \left[  \left[  1-p\left(  x\right)  \right]  +\left[  1-p\left(
y\right)  \right]  -\left[  1-p\left(  x,y\right)  \right]  \right]  $\\\hline
\multicolumn{1}{|r|}{$I(X,Y)=$} & $\sum_{x,y}p\left(  x,y\right)  \left[
\log\left(  \frac{1}{p\left(  x\right)  }\right)  +\log\left(  \frac
{1}{p\left(  y\right)  }\right)  -\log\left(  \frac{1}{p\left(  x,y\right)
}\right)  \right]  $\\\hline\hline
\multicolumn{1}{|r|}{$h\left(  p\Vert q\right)  =$} & $\frac{1}{2}\left[
\sum_{i}p_{i}(1-q_{i})+\sum_{i}q_{i}\left(  1-p_{i}\right)  \right]  $\\\hline
\multicolumn{1}{|r|}{$H_{s}(p||q)=$} & $\frac{1}{2}\left[  \sum_{i}p_{i}%
\log\left(  \frac{1}{q_{i}}\right)  +\sum_{i}q_{i}\log\left(  \frac{1}{p_{i}%
}\right)  \right]  $\\\hline\hline
\multicolumn{1}{|r|}{$d\left(  p||q\right)  =$} & $h\left(  p||q\right)
-\frac{1}{2}\left[  \left(  \sum_{i}p_{i}\left(  1-p_{i}\right)  \right)
+\left(  \sum_{i}q_{i}\left(  1-q_{i}\right)  \right)  \right]  $\\\hline
\multicolumn{1}{|r|}{$D_{s}\left(  p||q\right)  =$} & $H_{s}\left(
p||q\right)  -\frac{1}{2}\left[  \sum_{i}p_{i}\log\left(  \frac{1}{p_{i}%
}\right)  +\sum_{i}q_{i}\log\left(  \frac{1}{q_{i}}\right)  \right]
$\\\hline\hline
\end{tabular}

Table 4: The dit-bit transform from logical entropy to Shannon entropy
\end{center}

\section{Entropies for multivariate joint distributions}

Let $\left\{  p\left(  x_{1},...,x_{n}\right)  \right\}  $ be a probability
distribution on $X_{1}\times\ldots\times X_{n}$ for finite $X_{i}$'s. Let $S$
be a subset of $\left(  X_{1}\times...\times X_{n}\right)  ^{2}$ consisting of
certain ordered pairs of ordered $n$-tuples $\left(  \left(  x_{1}%
,...,x_{n}\right)  ,\left(  x_{1}^{\prime},...,x_{n}^{\prime}\right)  \right)
$ so the product probability measure on $S$ is:

\begin{center}
$\mu\left(  S\right)  =\sum\left\{  p\left(  x_{1},...,x_{n}\right)  p\left(
x_{1}^{\prime},...,x_{n}^{\prime}\right)  :\left(  \left(  x_{1}%
,...,x_{n}\right)  ,\left(  x_{1}^{\prime},...,x_{n}^{\prime}\right)  \right)
\in S\right\}  $.
\end{center}

\noindent Then all the logical entropies for this $n$-variable case are given
as the product measure of certain infosets $S$. Let $I,J\subseteq N$ be
subsets of the set of all variables $N=\left\{  X_{1},...,X_{n}\right\}  $ and
let $x=\left(  x_{1},...,x_{n}\right)  $ and similarly for $x^{\prime}$.

The joint logical entropy of all the variables is: $h\left(  X_{1}%
,...,X_{n}\right)  =\mu\left(  S_{\vee N}\right)  $ where:

\begin{center}
$S_{\vee N}=\left\{  \left(  x,x^{\prime}\right)  :%
%TCIMACRO{\tbigvee \limits_{i=1}^{n}}%
%BeginExpansion
{\textstyle\bigvee\limits_{i=1}^{n}}
%EndExpansion
x_{i}\neq x_{i}^{\prime}\right\}  =\cup\left\{  S_{X_{i}}:X_{i}\in N\right\}
$
\end{center}

\noindent(where $%
%TCIMACRO{\tbigvee }%
%BeginExpansion
{\textstyle\bigvee}
%EndExpansion
$ represents the disjunction of statements). For a non-empty $I\subseteq N$,
the joint logical entropy of the variables in $I$ could be represented as
$h\left(  I\right)  =\mu\left(  S_{\vee I}\right)  $ where:

\begin{center}
$S_{\vee I}=\left\{  \left(  x,x^{\prime}\right)  :%
%TCIMACRO{\tbigvee }%
%BeginExpansion
{\textstyle\bigvee}
%EndExpansion
x_{i}\neq x_{i}^{\prime}\text{ for }X_{i}\in I\right\}  =\cup\left\{
S_{X_{i}}:X_{i}\in I\right\}  $
\end{center}

\noindent so that $h\left(  X_{1},...,X_{n}\right)  =h\left(  N\right)  $.

As before, the information algebra $\mathcal{I}\left(  X_{1}\times...\times
X_{n}\right)  $ is the Boolean subalgebra of $\wp\left(  \left(  X_{1}%
\times...\times X_{n}\right)  ^{2}\right)  $ generated by the infosets
$S_{X_{i}}$ for the variables and their complements $S_{\lnot X_{i}}$.

For the conditional logical entropies, let $I,J\subseteq N$ be two non-empty
disjoint subsets of $N$. The idea for the conditional entropy $h\left(
I|J\right)  $ is to represent the information in the variables $I$ given by
the defining condition: $%
%TCIMACRO{\tbigvee }%
%BeginExpansion
{\textstyle\bigvee}
%EndExpansion
x_{i}\neq x_{i}^{\prime}$ for $X_{i}\in I$, after taking away the information
in the variables $J$ which is defined by the condition: $%
%TCIMACRO{\tbigvee }%
%BeginExpansion
{\textstyle\bigvee}
%EndExpansion
x_{j}\neq x_{j}^{\prime}$ for $X_{j}\in J$. Hence we negate that condition for
$J$ and add it to the condition for $I$ to obtain the conditional logical
entropy as $h\left(  I|J\right)  =\mu(S_{\vee I|\wedge J})$ where:

\begin{center}
$S_{\vee I|\wedge J}=\left\{  \left(  x,x^{\prime}\right)  :%
%TCIMACRO{\tbigvee }%
%BeginExpansion
{\textstyle\bigvee}
%EndExpansion
x_{i}\neq x_{i}^{\prime}\text{ for }X_{i}\in I\text{ and }%
%TCIMACRO{\tbigwedge }%
%BeginExpansion
{\textstyle\bigwedge}
%EndExpansion
x_{j}=x_{j}^{\prime}\text{ for }X_{j}\in J\right\}  $

$=\cup\left\{  S_{X_{i}}:X_{i}\in I\right\}  -\cup\left\{  S_{X_{j}}:X_{j}\in
J\right\}  =S_{\vee I}-S_{\vee J}$
\end{center}

\noindent(where $%
%TCIMACRO{\tbigwedge }%
%BeginExpansion
{\textstyle\bigwedge}
%EndExpansion
$ represents the conjunction of statements).

For the mutual logical information of a nonempty set of variables $I$,
$m\left(  I\right)  =\mu\left(  S_{\wedge I}\right)  $ where:

\begin{center}
$S_{\wedge I}=\left\{  \left(  x,x^{\prime}\right)  :%
%TCIMACRO{\tbigwedge }%
%BeginExpansion
{\textstyle\bigwedge}
%EndExpansion
x_{i}\neq x_{i}^{\prime}\text{ for }X_{i}\in I\right\}  $.
\end{center}

For the conditional mutual logical information, let $I,J\subseteq N$ be two
non-empty disjoint subsets of $N$ so that $m\left(  I|J\right)  =\mu\left(
S_{\wedge I|\wedge J}\right)  $ where:

\begin{center}
$S_{\wedge I|\wedge J}=\left\{  \left(  x,x^{\prime}\right)  :%
%TCIMACRO{\tbigwedge }%
%BeginExpansion
{\textstyle\bigwedge}
%EndExpansion
x_{i}\neq x_{i}^{\prime}\text{ for }X_{i}\in I\text{ and }%
%TCIMACRO{\tbigwedge }%
%BeginExpansion
{\textstyle\bigwedge}
%EndExpansion
x_{j}=x_{j}^{\prime}\text{ for }X_{j}\in J\right\}  $.
\end{center}

\noindent And finally by expressing the logical entropy formulas as averages,
the dit-bit transform will give the corresponding versions of Shannon entropy.

Consider an example of a joint distribution $\left\{  p\left(  x,y,z\right)
\right\}  $ on $X\times Y\times Z$. The mutual logical information $m\left(
X,Y,Z\right)  =\mu\left(  S_{\wedge\left\{  X,Y,Z\right\}  }\right)  $ where:

\begin{center}
$S_{\wedge\left\{  X,Y,Z\right\}  }=\left\{  \left(  \left(  x,y,z\right)
,\left(  x^{\prime},y^{\prime},z^{\prime}\right)  \right)  :x\neq x^{\prime
}\wedge y\neq y^{\prime}\wedge z\neq z^{\prime}\right\}  $.
\end{center}

%

%TCIMACRO{\FRAME{dtbpF}{2.2805in}{1.8697in}{0pt}{}{}{figure5.eps}%
%width 2.2805in;  height 1.8697in;  depth 0pt;  original-width 3.3892in;
%original-height 2.7743in;  cropleft "0";  croptop "1";  cropright "1";
%cropbottom "0";  filename 'figure5.eps';file-properties "XNPEU";}} }%
%BeginExpansion
\begin{center}
\includegraphics[
height=1.8697in,
width=2.2805in
]%
{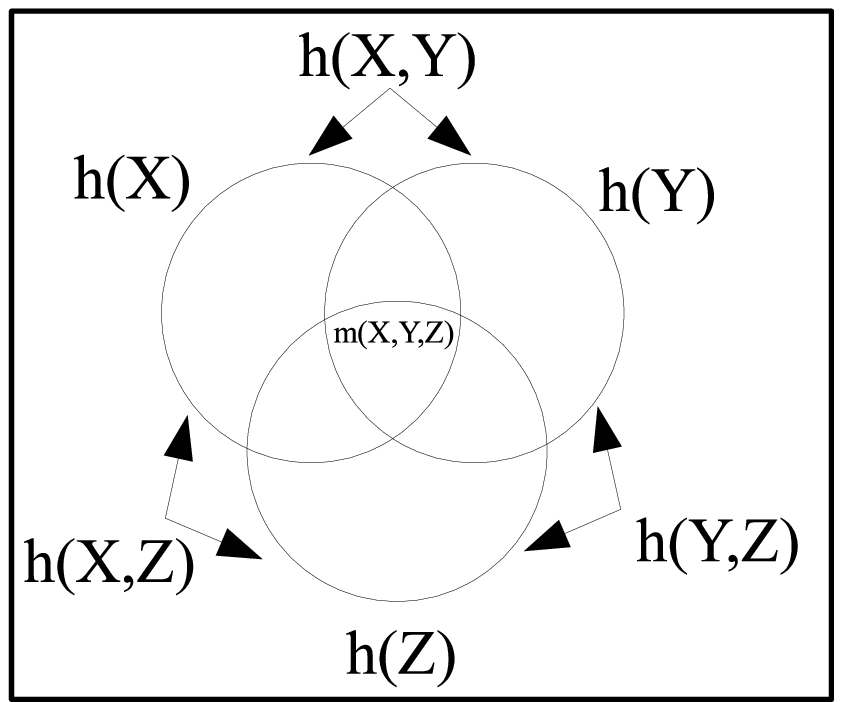}%
\end{center}
%EndExpansion

\begin{center}
Figure 5: Inclusion-exclusion principle for $h\left(  X,Y,Z\right)  $ (total
area within circles).
\end{center}

\noindent From the Venn diagram for $h\left(  X,Y,Z\right)  $, we have (using
a variation on the inclusion-exclusion principle)\footnote{The usual version
of the inclusion-exclusion principle would be: $h(X,Y,Z)=h(X)+h\left(
Y\right)  +h\left(  Z\right)  -m\left(  X,Y\right)  -m\left(  X,Z\right)
-m\left(  Y,Z\right)  +m\left(  X,Y,Z\right)  $ but $m\left(  X,Y\right)
=h(X)+h\left(  Y\right)  -h\left(  X,Y\right)  $ and so forth, so substituting
for $m\left(  X,Y\right)  $, $m\left(  X,Z\right)  $, and $m\left(
Y,Z\right)  $ gives the formula.}:

\begin{center}
$m\left(  X,Y,Z\right)  =h\left(  X\right)  +h\left(  Y\right)  +h\left(
Z\right)  -h\left(  X,Y\right)  -h\left(  X,Z\right)  -h\left(  Y,Z\right)
+h\left(  X,Y,Z\right)  $.
\end{center}

Substituting the averaging formulas for the logical entropies gives:

\begin{center}
$m\left(  X,Y,Z\right)  =$

$\sum_{x,y,z}p\left(  x,y,z\right)  \left[  \left[  1-p\left(  x\right)
\right]  +\left[  1-p\left(  y\right)  \right]  +\left[  1-p\left(  z\right)
\right]  -\left[  1-p\left(  x,y\right)  \right]  -\left[  1-p\left(
x,z\right)  \right]  -\left[  1-p\left(  y,z\right)  \right]  +\left[
1-p\left(  x,y,z\right)  \right]  \right]  $.
\end{center}

\noindent Then applying the dit-bit transform gives the corresponding formula
for the multivariate Shannon mutual information:\footnote{The multivariate
generalization of the Shannon mutual information was developed by William J.
McGill \cite{mcgill:psycho} and Robert M. Fano (\cite{fano:report149};
\cite{fano:trans}) at MIT in the early 50's and independently by Nelson M.
Blachman \cite{blachman:ire}. The criterion for it being the `correct'
generalization seems to be that it satisfied the generalized
inclusion-exclusion formulas (which generalize the two-variable Venn diagram)
that are automatically satisfied by any measure and are thus also obtained
\textit{from} the multivariate logical mutual information using the dit-bit
transform.}

\begin{center}
$I\left(  X,Y,Z\right)  =$

$\sum_{x,y,z}p\left(  x,y,z\right)  \left[  \log\left(  \frac{1}{p\left(
x\right)  }\right)  +\log\left(  \frac{1}{p\left(  y\right)  }\right)
+\log\left(  \frac{1}{p\left(  z\right)  }\right)  -\log\left(  \frac
{1}{p\left(  x,y\right)  }\right)  -\log\left(  \frac{1}{p\left(  x,z\right)
}\right)  -\log\left(  \frac{1}{p\left(  y,z\right)  }\right)  +\log\left(
\frac{1}{p\left(  x,y,z\right)  }\right)  \right]  $

$=\sum_{x,y,z}p\left(  x,y,z\right)  \left[  \log\left(  \frac{p\left(
x,y\right)  p\left(  x,z\right)  p\left(  y,z\right)  }{p\left(  x\right)
p\left(  y\right)  p\left(  z\right)  p\left(  x,y,z\right)  }\right)
\right]  $(e.g., \cite[p. 57]{fano:trans} or \cite[p. 129]{abramson:it}).
\end{center}

To emphasize that Venn-like diagrams are only a mnemonic analogy, Abramson
gives an example \cite[pp. 130-1]{abramson:it} where the Shannon mutual
information of three variables is negative.\footnote{Fano had earlier noted
that for three or more variables, the Shannon mutual information could be
negative. \cite[p. 58]{fano:trans}}

Consider the joint distribution $\left\{  p\left(  x,y,z\right)  \right\}  $on
$X\times Y\times Z$ where $X=Y=Z=\left\{  0,1\right\}  $.

\begin{center}%
\begin{tabular}
[c]{|c|c|c|c|c|c|}\hline
$X$ & $Y$ & $Z$ & $p(x,y,z)$ & $p(x,y),p(x,z),p\left(  y,z\right)  $ &
$p(x),p(y),p\left(  z\right)  $\\\hline\hline
$0$ & $0$ & $0$ & $\frac{1}{4}$ & $\frac{1}{4}$ & $\frac{1}{2}$\\\hline
$0$ & $0$ & $1$ & $0$ & $\frac{1}{4}$ & $\frac{1}{2}$\\\hline
$0$ & $1$ & $0$ & $0$ & $\frac{1}{4}$ & $\frac{1}{2}$\\\hline
$0$ & $1$ & $1$ & $\frac{1}{4}$ & $\frac{1}{4}$ & $\frac{1}{2}$\\\hline
$1$ & $0$ & $0$ & $0$ & $\frac{1}{4}$ & $\frac{1}{2}$\\\hline
$1$ & $0$ & $1$ & $\frac{1}{4}$ & $\frac{1}{4}$ & $\frac{1}{2}$\\\hline
$1$ & $1$ & $0$ & $\frac{1}{4}$ & $\frac{1}{4}$ & $\frac{1}{2}$\\\hline
$1$ & $1$ & $1$ & $0$ & $\frac{1}{4}$ & $\frac{1}{2}$\\\hline
\end{tabular}

Table 5: Abramson's example giving negative Shannon mutual information
$I\left(  X,Y,Z\right)  $.
\end{center}

Since the logical mutual information $m(X,Y,Z)$ is the measure $\mu\left(
S_{\wedge\left\{  X,Y,Z\right\}  }\right)  $, it is always non-negative and in
this case is $0$:

\begin{center}
$m\left(  X,Y,Z\right)  =h\left(  X\right)  +h\left(  Y\right)  +h\left(
Z\right)  -h\left(  X,Y\right)  -h\left(  X,Z\right)  -h\left(  Y,Z\right)
+h\left(  X,Y,Z\right)  $

$=\frac{1}{2}+\frac{1}{2}+\frac{1}{2}-\frac{3}{4}-\frac{3}{4}-\frac{3}%
{4}+\frac{3}{4}=\frac{3}{2}-\frac{6}{4}=0$.
\end{center}

\noindent All the compound notions of logical entropy have a direct
interpretation as a two-draw probability. The logical mutual information
$m\left(  X,Y,Z\right)  $ is the probability that in two independent samples
of $X\times Y\times Z$, the outcomes would differ in all coordinates. This
means the two draws would have the form $\left(  x,y,z\right)  $ and $\left(
1-x,1-y,1-z\right)  $ for the binary variables, but it is easily seen by
inspection that $p\left(  x,y,z\right)  =0$ or $p\left(  1-x,1-y,1-z\right)
=0$, so the products are $0$ as computed.

The Venn-diagram-like formula for $m(X,Y,Z)$ carries over to $I\left(
X,Y,Z\right)  $ by the dit-bit transform (since it preserves sums and
differences), but the \textquotedblleft area\textquotedblright\ $I(X,Y,Z)$ is negative:

\begin{center}
$I\left(  X,Y,Z\right)  =H\left(  X\right)  +H\left(  Y\right)  +H\left(
Z\right)  -H\left(  X,Y\right)  -H\left(  X,Z\right)  -H\left(  Y,Z\right)
+H\left(  X,Y,Z\right)  $

$=1+1+1-2-2-2+2=3-4=-1$.
\end{center}

\noindent It is unclear how that can be interpreted as the mutual information
contained in the three variables or how the corresponding \textquotedblleft
Venn diagram\textquotedblright\ (Figure 6) can be anything more than a
mnemonic for a formula. Indeed, as Csiszar and K\"{o}rner remark: 

\begin{quotation}
\noindent The set-function analogy might suggest to introduce further
information quantities corresponding to arbitrary Boolean expressions of sets.
E.g., the "information quantity" corresponding to $\mu\left(  A\cap B\cap
C\right)  =\mu\left(  A\cap B\right)  -\mu\left(  \left(  A\cap B\right)
-C\right)  $ would be $I(X,Y)-I(X,Y|Z)$; this quantity has, however, no
natural intuitive meaning. \cite[pp. 53-4]{csis-korn:infotheory}
\end{quotation}

Of course, all this works perfectly well in logical information theory for
\textquotedblleft arbitrary Boolean expressions of sets\textquotedblright\ in
the information algebra $\mathcal{I}\left(  X\times Y\times Z\right)  $, e.g.,
$m\left(  X,Y,Z\right)  =\mu\left(  S_{X}\cap S_{Y}\cap S_{Z}\right)
=\mu\left(  S_{X}\cap S_{Y}\right)  -\mu\left(  \left(  S_{X}\cap
S_{Y}\right)  -S_{Z}\right)  =m\left(  X,Y\right)  -m\left(  X,Y|Z\right)  $,
which also as a (two-draw) probability measure is always
non-negative.\footnote{The dit-bit transform turns the formula for $m\left(
X,Y,Z\right)  $ into the formula for $I\left(  X,Y,Z\right)  $, and it
preserves the Venn-diagram relationships but it does not preserve
non-negativity.}%

%TCIMACRO{\FRAME{dtbpF}{2.0064in}{1.772in}{0pt}{}{}{figure6.eps}%
%width 2.0064in;  height 1.772in;  depth 0pt;  original-width 2.5114in;
%original-height 2.2157in;  cropleft "0";  croptop "1";  cropright "1";
%cropbottom "0";  filename 'figure6.eps';file-properties "XNPEU";}} }%
%BeginExpansion
\begin{center}
\includegraphics[
height=1.772in,
width=2.0064in
]%
{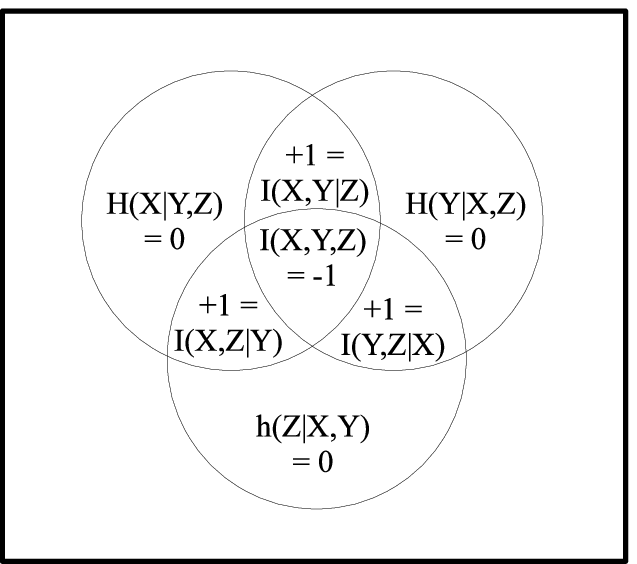}%
\end{center}
%EndExpansion

\begin{center}
Figure 6: Negative $I\left(  X,Y,Z\right)  $ in `Venn diagram.'
\end{center}

Note how the `intuitiveness' of independent random variables giving disjoint
Venn diagram circles comes back in a strange form in the multivariate case
since the three variables $X$, $Y$, and $Z$ in the example are pairwise
independent but not mutually independent (since any two determines the third).
Hence the circles for, say, $H\left(  X\right)  $ and $H\left(  Y\right)  $
`intersect' but the lense-shaped intersection is $I\left(  X,Y\right)
=I\left(  X,Y|Z\right)  +I\left(  X,Y,Z\right)  =+1-1=0$.

\section{Logical entropy as so-called `linear entropy'}

The Taylor series for $\ln(x+1)$ around $x=0$ is:

\begin{center}
$\ln(x+1)=\ln(1)+x-\frac{1}{2!}x^{2}+\frac{1}{3!}x^{3}2\left(  x+1\right)
^{-3}-...=x-\frac{x^{2}}{2}+\frac{x^{3}}{3}-...$
\end{center}

\noindent so substituting $x=p_{i}-1$ (with $p_{i}>0$) gives a version of the
Newton-Mercator series:

\begin{center}
$-\ln\left(  p_{i}\right)  =\ln\left(  \frac{1}{p_{i}}\right)  =1-p_{i}%
+\frac{\left(  p_{i}-1\right)  ^{2}}{2}-\frac{\left(  p_{i}-1\right)  ^{3}}%
{3}+....$
\end{center}

\noindent Then multiplying by $p_{i}$ and summing yields:

\begin{center}
$H_{e}\left(  p\right)  =-\sum_{i}p_{i}\ln\left(  p_{i}\right)  =\sum_{i}%
p_{i}\left(  1-p_{i}\right)  +\sum_{i}\frac{p_{i}(p_{i}-1)^{2}}{2}-...$

$=h\left(  p\right)  +\sum_{i}\frac{p_{i}(p_{i}-1)^{2}}{2}-...$.
\end{center}

\noindent A similar relationship holds in the quantum case between the von
Neumann entropy $S\left(  \rho\right)  =-\operatorname*{tr}\left[  \rho
\ln\left(  \rho\right)  \right]  $ and the \textit{quantum logical entropy}
$h\left(  \rho\right)  =\operatorname*{tr}\left[  \rho\left(  1-\rho\right)
\right]  =1-\operatorname*{tr}\left[  \rho^{2}\right]  $ which is defined by
having a density matrix $\rho$ replace the probability distribution $p$ and
the trace replace the sum.

Quantum logical entropy is beyond the scope of this paper but it might be
noted that some quantum information theorists have been using that concept to
rederive results previously derived using the von Neumann entropy such as the
Klein inequality, concavity, and a Holevo-type bound for Hilbert-Schmidt
distance (\cite{tamir-cohen:logicalentropy},
\cite{tamir-cohen:hilbert-schmidt}). There are many older results derived
under the misnomer \textquotedblleft linear entropy\textquotedblright\ or
derived for the quadratic special case of the Tsallis-Havrda-Charvat entropy
(\cite{havrda:alpha}; \cite{tsallis:entropy}, \cite{tsallis:nonext-sm}).
Moreover the logical derivation of the logical entropy formulas using the
notion of distinctions gives a certain naturalness to the notion of quantum
logical entropy.

\begin{quotation}
\noindent We find this framework of partitions and distinction most suitable
(at least conceptually) for describing the problems of quantum state
discrimination, quantum cryptography and in general, for discussing quantum
channel capacity. In these problems, we are basically interested in a distance
measure between such sets of states, and this is exactly the kind of knowledge
provided by logical entropy (\cite{ell:countingdits}). \cite[p. 1]%
{boaz-cohen:logicalentropy}
\end{quotation}

The relationship between the Shannon/von Neumann entropies and the logical
entropies in the classical and quantum cases is responsible for presenting the
logical entropy as a `linear' approximation to the Shannon or von Neumann
entropies since $1-p_{i}$ is the linear term in the series for $-\ln\left(
p_{i}\right)  $ [\textit{before} the multiplication by $p_{i}$ to make the
term quadratic!]. And $h\left(  p\right)  =1-\sum_{i}p_{i}^{2}$ or it quantum
counterpart $h\left(  \rho\right)  =1-\operatorname*{tr}\left[  \rho
^{2}\right]  $ are even called \textquotedblleft linear
entropy\textquotedblright\ (e.g., \cite{buscemi:lin-entropy} or
\cite{peters:linentropy}) even though the formulas are obviously
quadratic.\footnote{Sometimes the misnomer "linear entropy" is applied to the
rescaled logical entropy $\frac{n}{n-1}h\left(  \pi\right)  $. The maximum
value of the logical entropy is $h(\mathbf{1})=1-\frac{1}{n}=\frac{n-1}{n}$ so
the rescaling gives a maximum value of $1$. In terms of the partition-logic
derivation of the logical entropy formula, this amounts to sampling without
replacement and normalizing $\left\vert \operatorname*{dit}\left(  \pi\right)
\right\vert $ by the number of possible distinctions $\left\vert U\times
U-\Delta\right\vert =n^{2}-n$ (where $\Delta=\left\{  \left(  u,u\right)
:u\in U\right\}  $ is the diagonal) instead of $\left\vert U\times
U\right\vert =n^{2}$ since:
\par
\begin{center}
$\frac{\left\vert \operatorname*{dit}\left(  \pi\right)  \right\vert
}{\left\vert U\times U-\Delta\right\vert }=\frac{\left\vert
\operatorname*{dit}\left(  \pi\right)  \right\vert }{n\left(  n-1\right)
}=\frac{n}{n-1}\frac{\left\vert \operatorname*{dit}\left(  \pi\right)
\right\vert }{n^{2}}=\frac{n}{n-1}h\left(  \pi\right)  $.
\end{center}
} Another name for the quantum logical entropy found in the literature is
\textquotedblleft mixedness\textquotedblright\ \cite[p. 5]{jaeger:qinfo} which
at least doesn't call a quadratic formula `linear.' It is even called
\textquotedblleft impurity\textquotedblright\ since the complement $1-h\left(
\rho\right)  =\operatorname*{tr}\left[  \rho^{2}\right]  $ (i.e., the quantum
version of Turing's repeat rate $\sum_{i}p_{i}^{2}$) is called the
\textquotedblleft purity.\textquotedblright\ And as noted above, the formula
for logical entropy occurs as the quadratic special case of the
Tsallis-Havrda-Charvat entropy. Those parameterized families of entropy
formulas are sometimes criticized for lacking a convincing interpretation, but
we have seen that the quadratic case is based on partition logic dual to
Boole's subset logic. In terms of the duality between elements of a subset
(its) and distinctions of a partition (dits), the two measures are based on
the normalized counting measures of `its' and `dits').

In accordance with its quadratic nature, logical entropy is the logical
special case of C. R. Rao's quadratic entropy \cite{rao:div}. Two elements
from $U=\left\{  u_{1},...,u_{n}\right\}  $ are either identical or distinct.
Gini \cite{gini:vem} introduced $d_{ij}$ as the 'distance'\ between the
$i^{th}$ and $j^{th}$ elements where $d_{ij}=1$ for $i\not =j$ and $d_{ii}%
=0$--which might be considered the `logical distance function' $d_{ij}%
=1-\delta_{ij}$, the complement of the Kronecker delta. Since $1=\left(
p_{1}+...+p_{n}\right)  \left(  p_{1}+...+p_{n}\right)  =\sum_{i}p_{i}%
^{2}+\sum_{i\not =j}p_{i}p_{j}$, the logical entropy, i.e., Gini's index of
mutability, $h\left(  p\right)  =1-\sum_{i}p_{i}^{2}=\sum_{i\not =j}p_{i}%
p_{j}$, is the average logical distance between distinct elements. But one
might generalize by allowing other distances $d_{ij}=d_{ji}$ for $i\not =j$
(but always $d_{ii}=0$) so that $Q=\sum_{i\not =j}d_{ij}p_{i}p_{j}$ would be
the average distance between distinct elements from $U$. In 1982, C. R. Rao
introduced this concept as \textit{quadratic entropy} \cite{rao:div}.

Rao's treatment also includes (and generalizes) the natural extension of
logical entropy to continuous (square-integrable) probability density
functions $f\left(  x\right)  $ for a random variable $X$: $h\left(  X\right)
=1-\int f\left(  x\right)  ^{2}dx$. It might be noted that the natural
extension of Shannon entropy to continuous probability density functions
$f(x)$ through the limit of discrete approximations contains terms
$1/\log\left(  \Delta x_{i}\right)  $ that blow up as the mesh size $\Delta
x_{i}$ goes to zero (see \cite[pp. 34-38]{mceliece:info}).\footnote{For
expository purposes, we have restricted the treatment to finite sample spaces
$U$. For some countable discrete probability distributions, the Shannon
entropy blows up to infinity \cite[Example 2.46, p. 30]{yeung:firstcourse},
while the logical infosets are always well-defined and the logical entropy is
always in the half-open interval $[0,1)$.} Hence the definition of Shannon
entropy in the continuous case is defined not by the limit of the discrete
formula but by the \textit{analogous} formula $H\left(  X\right)  =-\int
f\left(  x\right)  \log\left(  f\left(  x\right)  \right)  dx$ which, as
McEliece points out, \textquotedblleft is not in any sense a measure of the
randomness of $X$\textquotedblright\ \cite[p. 38]{mceliece:info} in addition
to possibly having negative values. \cite[p. 74]{uffink:phd}

\section{On `intuitions' about information}

Lacking an immediate and convincing interpretation for an entropy formula, one
might produce a number of axioms about a `measure of information' where each
axiom is more or less intuitive. One supposed intuition about `information' is
that the information in independent random variables should be additive
(unlike probabilities $p\left(  x,y\right)  =p\left(  x\right)  p\left(
y\right)  $) or that the `information' in one variable conditional on a second
variable should be the same as the `information' in the first variable alone
when the variables are independent (like probabilities $p\left(  x|y\right)
=p\left(  x\right)  $).

Another intuition is that the information gathered from the occurrence of an
event is inversely related to the probability of the event. For instance, if
the probability of an outcome is $p_{i}$, then $\frac{1}{p_{i}}$ is a good
indicator of the surprise-value information gained by the occurrence of the
event. Very well; let us follow out that intuition to construct a
`surprise-value entropy.' We need to average the surprise-values across the
probability distribution $p=\left\{  p_{i}\right\}  =\left(  p_{1}%
,...,p_{n}\right)  $, and since the surprise-value is the multiplicative
inverse of the $p_{i}$, the natural notion of average is the multiplicative
(or geometric) average:

\begin{center}
$E\left(  p\right)  =\prod_{i=1}^{n}\left(  \frac{1}{p_{i}}\right)  ^{p_{i}}$.

Surprise-value entropy of a probability distribution $p=\left\{
p_{i}\right\}  =\left(  p_{1},...,p_{n}\right)  $.
\end{center}

How do the surprise-value intuitions square with intuitions about additive
information content for independent events? Given a joint probability
distribution $p_{xy}=p\left(  x,y\right)  $ on $X\times Y$, the two marginal
distributions are $p_{x}=\sum_{y}p_{xy}$ and $p_{y}=\sum_{x}p_{xy}$. Then we
showed previously that if the joint distribution was independent, i.e.,
$p_{xy}=p_{x}p_{y}$, then the Shannon entropies were additive (unlike probabilities):

\begin{center}
$H\left(  x,y\right)  =H\left(  x\right)  +H\left(  y\right)  $

Shannon entropies under independence.
\end{center}

\noindent This is in accordance with one `intuition' about independence.

But the surprise-value entropy is also based on intuitions so we need to check
if it is also additive for an independent joint distribution so that the
intuitions would be consistent. The surprise-value entropy of the independent
joint distribution $\left\{  p_{xy}\right\}  $ is:

\begin{center}
$E\left(  \left\{  p_{xy}\right\}  \right)  =\prod_{x,y}\left(  \frac
{1}{p_{xy}}\right)  ^{p_{xy}}=\prod_{x,y}\left(  \frac{1}{p_{x}p_{y}}\right)
^{p_{x}p_{y}}=\prod_{x}\prod_{y}\left(  \frac{1}{p_{x}}\right)  ^{p_{x}p_{y}%
}\left(  \frac{1}{p_{y}}\right)  ^{p_{x}p_{y}}$

$=\left[  \prod_{x}\prod_{y}\left(  \frac{1}{p_{x}}\right)  ^{p_{x}p_{y}%
}\right]  \left[  \prod_{y}\prod_{x}\left(  \frac{1}{p_{y}}\right)
^{p_{x}p_{y}}\right]  $

$=\left[  \prod_{x}\left(  \frac{1}{p_{x}}\right)  ^{p_{x}}\right]  \left[
\prod_{y}\left(  \frac{1}{p_{y}}\right)  ^{p_{y}}\right]  $

$=E\left(  \left\{  p_{x}\right\}  \right)  E\left(  \left\{  p_{y}\right\}
\right)  $
\end{center}

\noindent so the surprise-value of an independent joint distribution is the
\textit{product} of the surprise-value entropies of the marginal distributions
(like probabilities). The derivation used the fact that the multiplicative
average of a constant is, of course, that constant, e.g., $\prod_{y}c^{p_{y}%
}=c^{\sum_{y}p_{y}}=c$.

Since the two intuitions give conflicting results, which, if either, is
`correct'? Should `entropy' be additive or multiplicative for independent
distributions? At this point, it is helpful to step back and note that in
statistics, for example, any product of random variables $XY$ can sometimes,
with advantage, be analyzed using the sum of log-variables, $\log\left(
XY\right)  =\log\left(  X\right)  +\log\left(  Y\right)  $. It is best seen as
a question of convenience rather than `truth' whether to use the product or
the log of the product.

In the case at hand, the notion of surprise-value entropy, which is
multiplicative for independent distributions, can trivially be turned into an
expression that is additive for independent distributions by taking logarithms
to some base:

\begin{center}
$\log E\left(  \left[  p_{xy}\right]  \right)  =\log E\left(  \left[
p_{x}\right]  \right)  +\log E\left(  \left[  p_{y}\right]  \right)  $.
\end{center}

\noindent Is the original surprise-value formula $E\left(  p\right)  $ or the
log-of-surprise-value formula $\log E\left(  p\right)  $ the `true' measure?
And, in the case at hand, the point is that the log-of-surprise-value formula
\textit{is} the Shannon entropy:

\begin{center}
$\log E\left(  p\right)  =H\left(  p\right)  $ or $E\left(  p\right)
=2^{H\left(  p\right)  }$.
\end{center}

Some authors have even suggested that the surprise-value formula is
\textit{more} intuitive than the log-formula. To understand this intuition, we
need to develop another interpretation of the surprise-value formula. When an
event or outcome has a probability $p_{i}$, it is intuitive to think of it as
being drawn from a set of $\frac{1}{p_{i}}$ equiprobable elements
(particularly when $\frac{1}{p_{i}}$ is an integer) so $\frac{1}{p_{i}}$ is
called the \textit{numbers-equivalent} \cite{adel:ne} of the probability
$p_{i}$. Hence the multiplicative average of the numbers-equivalents for a
probability distribution $p=\left(  p_{1},...,p_{n}\right)  $ is $E\left(
p\right)  $, which thus will now be called the \textit{numbers-equivalent
entropy }(also called \textquotedblleft exponential entropy\textquotedblright%
\ \cite{camp:exp}). This approach also supplies an interpretation: Sampling a
probability distribution $p$ is like, on average, sampling from a distribution
with $E\left(  p\right)  $ equiprobable outcomes.

In the biodiversity literature, the situation is that each animal (in a
certain territory) is considered to be equiprobable to be sampled and the
partition of the animals is by species. Taking $p=\left(  p_{1},...,p_{n}%
\right)  $ as the probability distribution of the $n$ species, the
numbers-equivalent entropy $E\left(  p\right)  $ is the measure of
biodiversity that says sampling the population is like sampling a population
of $E\left(  p\right)  $ equally common species. The mathematical biologist
Robert H. MacArthur finds this much more intuitive than Shannon entropy.

\begin{quotation}
\noindent Returning to the example of a census of $99$ individuals of one
species and $1$ of a second, we calculate $H=...=0.0560$ [as the Shannon
entropy using natural logs]. For a census of fifty individuals of each of the
two species we would get $H=...=0.693$. To convert these back to `equally
common species', we take $e^{0.0560}=1.057$ for the first census and
$e^{0.693}=2.000$ for the second. These numbers, $1.057$ and $2$, accord much
more closely with our intuition of how diverse the areas actually are,... .
\cite[p. 514]{macarthur:div}
\end{quotation}

\noindent MacArthur's interpretation is \textquotedblleft that [$E\left(
p\right)  $] equally common species would have the same diversity as the [$n$]
unequally common species in our census.\textquotedblright\ \cite[p.
514]{macarthur:div}

The point is that `intuitions' differ even between Shannon entropy $H\left(
p\right)  $ and its base-free anti-log $E\left(  p\right)  $, not to mention
between other approaches to entropy. There is now in the literature a
`veritable plethora' of entropy definitions (\cite{aczel-d:measures};
\cite{tsallis:nonext-sm}) each with its `intuitive axioms.' Surely there are
better criteria for entropy concepts that differing subjective intuitions. 

\section{The connection with entropy in statistical mechanics}

Shannon entropy is sometimes referred to as \textquotedblleft
Boltzmann-Shannon entropy\textquotedblright\ or \textquotedblleft
Boltzmann-Gibbs-Shannon entropy\textquotedblright\ since the Shannon formula
supposedly has the same functional form as Boltzmann entropy which even
motivated the name \textquotedblleft entropy.\textquotedblright\ The name
\textquotedblleft entropy\textquotedblright\ is here to stay, but the
justification of the formula by reference to statistical mechanics is not
quite correct.

The connection between entropy in statistical mechanics and Shannon's entropy
is only via a numerical approximation, the Stirling approximation, where if
the first two terms in the Stirling approximation are used, then the Shannon
formula is obtained. The first two terms in the Stirling approximation for
$\ln(N!)$ are: $\ln\left(  N!\right)  \approx N\ln(N)-N$. The first three
terms in the Stirling approximation are: $\ln\left(  N!\right)  \approx
N(\ln(N)-1)+\frac{1}{2}\ln\left(  2\pi N\right)  $.

If we consider a partition on a finite $U$ with $\left\vert U\right\vert =N$,
with $n$ blocks of size $N_{1},...,N_{n}$, then the number of ways of
distributing the individuals in these $n$ boxes with those numbers $N_{i}$ in
the $i^{th}$ box is: $W=\frac{N!}{N_{1}!...N_{n}!}$. The normalized natural
log of $W$, $S=\frac{1}{N}\ln\left(  W\right)  $ is one form of entropy in
statistical mechanics. Indeed, the formula \textquotedblleft$S=k\log\left(
W\right)  $\textquotedblright\ is engraved on Boltzmann's tombstone.

The entropy formula:

\begin{center}
$S=\frac{1}{N}\ln\left(  W\right)  =\frac{1}{N}\ln\left(  \frac{N!}%
{N_{1}!...N_{n}!}\right)  =\frac{1}{N}\left[  \ln(N!)-\sum_{i}\ln
(N_{i}!)\right]  $
\end{center}

\noindent can then be developed using the first two terms in the Stirling approximation

\begin{center}
$\frac{1}{N}\ln\left(  W\right)  \approx\frac{1}{N}\left[  N\left[  \ln\left(
N\right)  -1\right]  -\sum_{i}N_{i}\left[  \ln\left(  N_{i}\right)  -1\right]
\right]  $

$=\frac{1}{N}\left[  N\ln(N)-\sum N_{i}\ln(N_{i})\right]  =\frac{1}{N}\left[
\sum N_{i}\ln\left(  N\right)  -\sum N_{i}\ln\left(  N_{i}\right)  \right]  $

$=\sum\frac{N_{i}}{N}\ln\left(  \frac{1}{N_{i}/N}\right)  =\sum p_{i}%
\ln\left(  \frac{1}{p_{i}}\right)  =H_{e}\left(  p\right)  $
\end{center}

\noindent where $p_{i}=\frac{N_{i}}{N}$ (and where the formula with logs to
the base $e$ only differs from the usual base $2$ formula by a scaling
factor). Shannon's entropy $H_{e}\left(  p\right)  $ is in fact an excellent
numerical approximation to Boltzmann entropy $S=\frac{1}{N}\ln\left(
W\right)  $ for large $N$ (e.g., in statistical mechanics). But that does not
justify using expressions like \textquotedblleft Boltzmann-Shannon
entropy\textquotedblright\ as if the log of the combinatorial formula $W$
involving factorials was the same as the two-term Stirling approximation.

The common claim that Shannon's entropy has the \textit{same functional form}
as entropy in statistical mechanics is simply false. If we use a three-term
Stirling approximation, then we obtain an \textit{even better} numerical
approximation:\footnote{For the case $n=2$, MacKay \cite[p. 2]{mackay:info}
also uses the next term in the Stirling's approximation to give a "more
accurate approximation" to the entropy of statistical mechanics than the
Shannon entropy (the two-term approximation).}

\begin{center}
$S=\frac{1}{N}\ln\left(  W\right)  \approx H_{e}\left(  p\right)  +\frac
{1}{2N}\ln\left(  \frac{2\pi N^{n}}{\left(  2\pi\right)  ^{n}\Pi p_{i}%
}\right)  $
\end{center}

\noindent but no one would suggest using that \textquotedblleft more
accurate\textquotedblright\ entropy formula in information theory or dream of
calling it the \textquotedblleft Boltzmann-Shannon entropy.\textquotedblright%
\ Shannon's formula should be justified and understood on its own terms (see
next section), and not by over-interpreting the numerically approximate
relationship with entropy in statistical mechanics.

\section{The statistical interpretation of Shannon entropy}

Shannon, like Ralph Hartley \cite{hart:ti} before him, starts with the
question of how much `information' is required to single out a designated
element from a set $U$ of equiprobable elements. Renyi formulated this in
terms of the search \cite{Renyi:pt} for a hidden element like the answer in a
Twenty Questions game or the sent message in a communication. But being able
to always find the designated element is equivalent to being able to
distinguish all elements from one another.

One might quantify `information' as the minimum number of yes-or-no questions
in a game of Twenty Questions that it would take in general to
\textit{distinguish} all the possible \textquotedblleft
answers\textquotedblright\ (or \textquotedblleft messages\textquotedblright%
\ in the context of communications). This is readily seen in the simple case
where $\left\vert U\right\vert =n=2^{m}$, i.e., the size of the set of
equiprobable elements is a power of $2$. Then following the lead of Wilkins
over three centuries earlier, the $2^{m}$ elements could be encoded using
words of length $m$ in a binary code such as the digits $\left\{  0,1\right\}
$ of binary arithmetic (or $\left\{  A,B\right\}  $ in the case of Wilkins).
Then an efficient or minimum set of yes-or-no questions needed to single out
the hidden element is the set of $m$ questions:

\begin{center}
\textquotedblleft Is the $j^{th}$ digit in the binary code for the hidden
element a $1$?\textquotedblright
\end{center}

\noindent for $j=1,...,m$. Each element is distinguished from any other
element by their binary codes differing in at least one digit. The information
gained in finding the outcome of an equiprobable binary trial, like flipping a
fair coin, is what Shannon calls a \textit{bit} (derived from
\textquotedblleft binary digit\textquotedblright). Hence the information
gained in distinguishing all the elements out of $2^{m}$ equiprobable elements is:

\begin{center}
$m=\log_{2}\left(  2^{m}\right)  =\log_{2}\left(  \left\vert U\right\vert
\right)  =\log_{2}\left(  \frac{1}{p_{0}}\right)  $ bits
\end{center}

\noindent where $p_{0}=\frac{1}{2^{m}}$ is the probability of any given
element (henceforth all logs to base $2$).\footnote{This is the special case
where Campbell \cite{camp:meas} noted that Shannon entropy acted as a measure
to count that number of binary partitions.}

In the more general case where $\left\vert U\right\vert =n$ is not a power of
$2$, Shannon and Hartley extrapolate to the definition of $H\left(
p_{0}\right)  $ where $p_{0}=\frac{1}{n}$ as:

\begin{center}
$H\left(  p_{0}\right)  =\log\left(  \frac{1}{p_{0}}\right)  =\log\left(
n\right)  $

Shannon-Hartley entropy for an equiprobable set $U$ of $n$ elements.
\end{center}

\noindent The Shannon formula then extrapolates further to the case of
different probabilities $p=\left(  p_{1},...,p_{n}\right)  $ by taking the average:

\begin{center}
$H\left(  p\right)  =\sum_{i=1}^{n}p_{i}\log_{2}\left(  \frac{1}{p_{i}%
}\right)  $.

Shannon entropy for a probability distribution $p=\left(  p_{1},...,p_{n}%
\right)  $
\end{center}

How can that extrapolation and averaging be made rigorous to offer a more
convincing interpretation? Shannon uses the law of large numbers. Suppose that
we have a three-letter alphabet $\left\{  a,b,c\right\}  $ where each letter
was equiprobable, $p_{a}=p_{b}=p_{c}=\frac{1}{3}$, in a multi-letter message.
Then a one-letter or two-letter message cannot be exactly coded with a binary
$0,1$ code with equiprobable $0$'s and $1$'s. But any probability can be
better and better approximated by longer and longer representations in the
binary number system. Hence we can consider longer and longer messages of $N$
letters along with better and better approximations with binary codes. The
long run behavior of messages $u_{1}u_{2}...u_{N}$ where $u_{i}\in\left\{
a,b,c\right\}  $ is modeled by the law of large numbers so that the letter $a$
on average occur $p_{a}N=\frac{1}{3}N$ times and similarly for $b$ and $c$.
Such a message is called \textit{typical}.

The probability of any one of those typical messages is:

\begin{center}
$p_{a}^{p_{a}N}p_{b}^{p_{b}N}p_{c}^{p_{c}N}=\left[  p_{a}^{p_{a}}p_{b}^{p_{b}%
}p_{c}^{p_{c}}\right]  ^{N}$
\end{center}

or, in this case,

\begin{center}
$\left[  \left(  \frac{1}{3}\right)  ^{1/3}\left(  \frac{1}{3}\right)
^{1/3}\left(  \frac{1}{3}\right)  ^{1/3}\right]  ^{N}=\left(  \frac{1}%
{3}\right)  ^{N}$.
\end{center}

\noindent Hence the number of such typical messages is $3^{N}$.

If each message was assigned a unique binary code, then the number of $0,1$'s
in the code would have to be $X$ where $2^{X}=3^{N}$ or $X=\log\left(
3^{N}\right)  =N\log\left(  3\right)  $. Hence the number of equiprobable
binary questions or bits needed per letter (i.e., to distinguish each letter)
of a typical message is:

\begin{center}
$N\log(3)/N=\log\left(  3\right)  =3\times\frac{1}{3}\log\left(  \frac{1}%
{1/3}\right)  =H\left(  p\right)  $.
\end{center}

\noindent This example shows the general pattern.

In the general case, let $p=\left(  p_{1},...,p_{n}\right)  $ be the
probabilities over a $n$-letter alphabet $A=\left\{  a_{1},...,a_{n}\right\}
$. In an $N$-letter message, the probability of a particular message
$u_{1}u_{2}...u_{N}$ is $\Pi_{i=1}^{N}\Pr\left(  u_{i}\right)  $ where $u_{i}$
could be any of the symbols in the alphabet so if $u_{i}=a_{j}$ then
$\Pr\left(  u_{i}\right)  =p_{j}$.

In a \textit{typical} message, the $i^{th}$ symbol will occur $p_{i}N$ times
(law of large numbers) so the probability of a typical message is (note change
of indices to the letters of the alphabet):

\begin{center}
$\Pi_{k=1}^{n}p_{k}^{p_{k}N}=\left[  \Pi_{k=1}^{n}p_{k}^{p_{k}}\right]  ^{N}$.
\end{center}

Thus the probability of a typical message is $P^{N}$ where it is as if each
letter in a typical message was equiprobable with probability $P=\Pi_{k=1}%
^{n}p_{k}^{p_{k}}$. No logs have been introduced into the argument yet, so we
have an interpretation of the base-free numbers-equivalent entropy $E\left(
p\right)  =P^{-1}$: it is as if each letter in a typical message is being draw
from an alphabet with $P^{-1}=\Pi_{k=1}^{n}p_{k}^{-p_{k}}$ equiprobable
letters. Hence the number of $N$-letter messages from the equiprobable
alphabet is then $\left[  \Pi_{k=1}^{n}p_{k}^{-p_{k}}\right]  ^{N}$. The
choice of base $2$ means assigning a unique binary code to each typical
message requires $X$ bits where $2^{X}=\left[  \Pi_{k=1}^{n}p_{k}^{-p_{k}%
}\right]  ^{N}$ where:

\begin{center}
$X=\log\left\{  \left[  \Pi_{k=1}^{n}p_{k}^{-p_{k}}\right]  ^{N}\right\}
=N\log\left[  \Pi_{k=1}^{n}p_{k}^{-p_{k}}\right]  $

$=N\sum_{k=1}^{n}\log\left(  p_{k}^{-p_{k}}\right)  =N\sum_{k}-p_{k}%
\log\left(  p_{k}\right)  $

$=N\sum_{k}p_{k}\log\left(  \frac{1}{p_{k}}\right)  =NH\left(  p\right)  $.
\end{center}

\noindent Dividing by the number $N$ of letters gives the average bit-count
interpretation of the Shannon entropy; $H\left(  p\right)  =\sum_{k=1}%
^{n}p_{k}\log\left(  \frac{1}{p_{k}}\right)  $ is the \textit{average number
of bits necessary to distinguish each letter in a typical message}.

This result, usually called the \textit{noiseless coding theorem}, allows us
to \textit{conceptually} relate the logical and Shannon entropies (the dit-bit
transform gives the quantitative relationship). In terms of the simplest case
for partitions, the Shannon entropy $H\left(  \pi\right)  =\sum_{B\in\pi}%
p_{B}\log_{2}\left(  1/p_{B}\right)  =-\sum_{B\in\pi}p_{B}\log_{2}\left(
p_{B}\right)  $ is a \textit{requantification} of the logical measure of
information $h\left(  \pi\right)  =\frac{|\operatorname*{dit}\left(
\pi\right)  |}{\left\vert U\times U\right\vert }=1-\sum_{B\in\pi}p_{B}^{2}$.
Instead of directly counting the distinctions of $\pi$, the idea behind
Shannon entropy is to count the (minimum) number of binary partitions needed
to make all the distinctions of $\pi$. In the special case of $\pi$ having
$2^{m}$ equiprobable blocks, the number of binary partitions $\beta_{i}$
needed to make the distinctions $\operatorname*{dit}\left(  \pi\right)  $ of
$\pi$ is $m$. Represent each block by an $m$-digit binary number so the
$i^{th}$ binary partition $\beta_{i}$ just distinguishes those blocks with
$i^{th}$ digit $0$ from those with $i^{th}$ digit $1$.\footnote{Thus as noted
by John Wilkins in 1641, five letter words in a two-letter code would suffice
to distinguish $2^{5}=32$ distinct entities. \cite{wilkins:merc}} Thus there
are $m$ binary partitions $\beta_{i}$ such that $%
%TCIMACRO{\tbigvee \limits_{i=1}^{m}}%
%BeginExpansion
{\textstyle\bigvee\limits_{i=1}^{m}}
%EndExpansion
\beta_{i}=\pi$ or, equivalently, $%
%TCIMACRO{\tbigcup \limits_{i=1}^{m}}%
%BeginExpansion
{\textstyle\bigcup\limits_{i=1}^{m}}
%EndExpansion
\operatorname*{dit}\left(  \beta_{i}\right)  =\operatorname*{dit}\left(
%TCIMACRO{\tbigvee \limits_{i=1}^{m}}%
%BeginExpansion
{\textstyle\bigvee\limits_{i=1}^{m}}
%EndExpansion
\beta_{i}\right)  =\operatorname*{dit}\left(  \pi\right)  $. Thus $m$ is the
exact number of binary partitions it takes to make the distinctions of $\pi$.
In the general case, Shannon gives the above statistical interpretation so
that $H\left(  \pi\right)  $ is the minimum \textit{average} number of binary
partitions or bits needed to make the distinctions of $\pi$.

Note the difference in emphasis. Logical information theory is only concerned
with counting the distinctions between distinct elements, not with uniquely
designating the distinct entities. By requantifying to count the number of
binary partitions it takes to make the same distinctions, the emphasis shifts
to the length of the binary code necessary to uniquely designate the distinct
elements. Thus the Shannon information theory perfectly dovetails into coding
theory and is often presented today as the unified theory of information and
coding (e.g., \cite{mceliece:info} or \cite{hamming:codinginfo}). It is that
shift to not only making distinctions but uniquely coding the distinct
outcomes that gives the Shannon theory of information, coding, and
communication such importance in applications.

\section{Concluding remarks}

The answer to the title question is that partition logic gives a derivation of
the (old) formula $h\left(  \pi\right)  =1-\sum_{i}p_{B_{i}}^{2}$ for
partitions as the normalized counting measure on the distinctions (`dits') of
a partition $\pi=\left(  B_{1},...,B_{m}\right)  $ that is the analogue of the
Boolean subset logic derivation of logical probability as the normalized
counting measure on the elements (`its') of a subset. In short, logical
information is the quantitative measure built on top of partition logic just
as logical probability is the quantitative measure built on top of ordinary
subset logic which might be symbolized as:

\begin{center}
$\frac{logical\text{ }information}{partition\text{ }logic}=\frac{logical\text{
}probability}{subset\text{ }logic}$.
\end{center}

\noindent Since conventional information theory has heretofore been focused on
the original notion of Shannon entropy (and quantum information theory on the
corresponding notion of von Neumann entropy), much of the paper has compared
the logical entropy notions to the corresponding Shannon entropy notions.

Logical entropy, like logical probability, is a measure, while Shannon entropy
is not. The compound Shannon entropy concepts nevertheless satisfy the
measure-like Venn diagram relationships that are automatically satisfied by a
measure. This can be explained by the dit-bit transform so that by putting a
logical entropy notion into the proper form as an average of dit-counts, one
can replace a dit-count by a bit-count and obtain the corresponding Shannon
entropy notion--which shows why the latter concepts satisfy the same Venn
diagram relationships.

Other comparisons were made in terms of the various `intuitions' expressed in
axioms, on the alleged identity in functional form between Shannon entropy and
entropy in statistical mechanics, and on the statistical interpretation of
Shannon entropy and its base-free antilog, the numbers-equivalent entropy
$E\left(  p\right)  =2^{H\left(  p\right)  }$.

The basic idea of information is distinctions, and distinctions have a precise
definition (dits) in partition logic. Prior to using any probabilities,
logical information theory defines the information sets (i.e., sets of
distinctions) which for partitions are the ditsets. Given a probability
distribution on a set $U$, the product probability measure on $U\times U$
applied to the information sets gives the quantitative notion of logical
entropy. Information sets and logical entropy give the basic combinatorial and
quantitative notions of information-as-distinctions. Shannon entropy is a
requantification (well-adapted for the theory of coding and communication)
that counts the minimum number of binary partitions (bits) that are required,
on average, to make all the same distinctions, i.e., to encode the
distinguished elements.


\begin{thebibliography}{99}                                                                                               %


\bibitem {abramson:it}Abramson, Norman 1963. \textit{Information Theory and
Coding}. New York: McGraw-Hill.

\bibitem {aczel-d:measures}Aczel, J., and Z. Daroczy. 1975. \textit{On
Measures of Information and Their Characterization}. New York: Academic Press.

\bibitem {adel:ne}Adelman, M. A. 1969. Comment on the H Concentration Measure
as a Numbers-Equivalent. \textit{Review of Economics and Statistics}. 51: 99-101.

\bibitem {adriaans-benthem:philinfo}Adriaans, Pieter, and Johan van Benthem,
eds. 2008. \textit{Philosophy of Information. Vol. 8. Handbook of the
Philosophy of Science}. Amsterdam: North-Holland.

\bibitem {bennett:qinfo}Bennett, Charles H. 2003. Quantum Information: Qubits
and Quantum Error Correction. \textit{International Journal of Theoretical
Physics} 42 (2 February): 153--76.

\bibitem {blachman:ire}Blachman, Nelson M. 1961. A Generalization of Mutual
Information. \textit{Proc. IRE} 49 (8 August): 1331--32.

\bibitem {boole:lot}Boole, George 1854. \textit{An Investigation of the Laws
of Thought on which are founded the Mathematical Theories of Logic and
Probabilities}. Cambridge: Macmillan and Co.

\bibitem {buscemi:lin-entropy}Buscemi, Fabrizio, Paolo Bordone, and Andrea
Bertoni. 2007. Linear Entropy as an Entanglement Measure in Two-Fermion
Systems. \textit{ArXiv.org}. March 2. http://arxiv.org/abs/quant-ph/0611223v2.

\bibitem {camp:meas}Campbell, L. Lorne 1965. Entropy as a
Measure.\textit{\ IEEE Trans. on Information Theory}. IT-11 (January): 112-114.

\bibitem {camp:exp}Campbell, L. Lorne. 1966. Exponential Entropy as a Measure
of the Extent of a Distribution. \textit{Zeitschrift F\"{u}r
Wahrscheinlichkeitstheorie Und Verwandte Gebiete} 5: 217--25.

\bibitem {cover:eit}Cover, Thomas and Joy Thomas 1991. \textit{Elements of
Information Theory}. New York: John Wiley.

\bibitem {csis-korn:infotheory}Csiszar, Imre, and Janos K\"{o}rner. 1981.
\textit{Information Theory: Coding Theorems for Discrete Memoryless Systems}.
New York: Academic Press.

\bibitem {ell:countingdits}Ellerman, David. 2009. Counting Distinctions: On
the Conceptual Foundations of Shannon's Information Theory. \textit{Synthese}
168 (1 May): 119--49.

\bibitem {ell:partitions}Ellerman, David 2010. The Logic of Partitions:
Introduction to the Dual of the Logic of Subsets. \textit{Review of Symbolic
Logic}. 3 (2 June): 287-350.

\bibitem {ell:intropartlogic}Ellerman, David 2014. An Introduction of
Partition Logic. \textit{Logic Journal of the IGPL.} 22, no. 1: 94--125.

\bibitem {fano:report149}Fano, Robert M. 1950. The Transmission of Information
II. \textit{Research Laboratory of Electronics Report 149}. Cambridge MA: MIT.

\bibitem {fano:trans}Fano, Robert M. 1961. \textit{Transmission of
Information}. Cambridge MA: MIT Press.

\bibitem {fried:ioc}Friedman, William F. 1922. \textit{The Index of
Coincidence and Its Applications in Cryptography}. Geneva IL: Riverbank Laboratories.

\bibitem {gini:vem}Gini, Corrado 1912. \textit{Variabilit\`{a} e
mutabilit\`{a}}. Bologna: Tipografia di Paolo Cuppini.

\bibitem {gini:vemrpt}Gini, Corrado 1955. Variabilit\`{a} e mutabilit\`{a}. In
\textit{Memorie di metodologica statistica}. E. Pizetti and T. Salvemini eds.,
Rome: Libreria Eredi Virgilio Veschi.

\bibitem {gleick:info}Gleick, James 2011. \textit{The Information: A History,
A Theory, A Flood}. New York: Pantheon.

\bibitem {good:turing}Good, I. J. 1979. A.M. Turing's statistical work in
World War II. \textit{Biometrika}. 66 (2): 393-6.

\bibitem {good:div}Good, I. J. 1982. Comment (on Patil and Taillie: Diversity
as a Concept and its Measurement). \textit{Journal of the American Statistical
Association}. 77 (379): 561-3.

\bibitem {hamming:codinginfo}Hamming, Richard W. 1980. \textit{Coding and
Information Theory}. Englewood Cliffs, NJ: Prentice-Hall.

\bibitem {hart:ti}Hartley, Ralph V. L. 1928. Transmission of information.
\textit{Bell System Technical Journal}. 7 (3, July): 535-63.

\bibitem {havrda:alpha}Havrda, Jan, and Frantisek Charvat. 1967.
Quantification Methods of Classification Processes: Concept of Structural
$\alpha$-Entropy. \textit{Kybernetika} (Prague) 3: 30--35.

\bibitem {her:conc}Herfindahl, Orris C. 1950. \textit{Concentration in the
U.S. Steel Industry}. Unpublished doctoral dissertation, Columbia University.

\bibitem {hirsch:pat}Hirschman, Albert O. 1964. The Paternity of an Index.
\textit{American Economic Review}. 54 (5): 761-2.

\bibitem {hu:info}Hu, Guo Ding. 1962. On the Amount of Information (in
Russian). \textit{Teor. Veroyatnost. I Primenen}. 4: 447--55.

\bibitem {jaeger:qinfo}Jaeger, Gregg. 2007. \textit{Quantum Information: An
Overview}. New York: Springer Science+Business Media.

\bibitem {kolmogor:combfound}Kolmogorov, A. N. 1983. Combinatorial Foundations
of Information Theory and the Calculus of Probabilities. \textit{Russian Math.
Surveys} 38 (4): 29--40.

\bibitem {kull:crypt}Kullback, Solomon 1976. \textit{Statistical Methods in
Cryptanalysis}. Walnut Creek CA: Aegean Park Press.

\bibitem {laplace:probs}Laplace, Pierre-Simon. 1995 (1825).
\textit{Philosophical Essay on Probabilities}. Translated by A. I. Dale. New
York: Springer Verlag.

\bibitem {law:sfm}Lawvere, F. William and Robert Rosebrugh 2003. \textit{Sets
for Mathematics}. Cambridge: Cambridge University Press.

\bibitem {macarthur:div}MacArthur, Robert H. 1965. Patterns of Species
Diversity. \textit{Biol. Rev.} 40: 510--33.

\bibitem {mackay:info}MacKay, David J. C. 2003. \textit{Information Theory,
Inference, and Learning Algorithms}. Cambridge UK: Cambridge University Press.

\bibitem {mcgill:psycho}McGill, William J. 1954. Multivariate Information
Transmission. \textit{Psychometrika} 19 (2 June): 97--116.

\bibitem {mceliece:info}McEliece, R. J. 1977. \textit{The Theory of
Information and Coding: A Mathematical Framework for Communication
(Encyclopedia of Mathematics and Its Applications, Vol. 3)}. Reading MA: Addison-Wesley.

\bibitem {nielsen-chuang:bible}Nielsen, M., and I. Chuang. 2000.
\textit{Quantum Computation and Quantum Information}. Cambridge: Cambridge
University Press.

\bibitem {patil:div}Patil, G. P. and C. Taillie 1982. Diversity as a Concept
and its Measurement. \textit{Journal of the American Statistical Association}.
77 (379): 548-61.

\bibitem {peters:linentropy}Peters, Nicholas A., Tzu-Chieh Wei, and Paul G.
Kwiat. 2004. Mixed State Sensitivity of Several Quantum Information
Benchmarks. \textit{ArXiv.org}. October 22. http://arxiv.org/abs/quant-ph/0407172v2.

\bibitem {rao:div}Rao, C. R. 1982. Diversity and Dissimilarity Coefficients: A
Unified Approach. \textit{Theoretical Population Biology}. 21: 24-43.

\bibitem {Renyi:pt}R\'{e}nyi, Alfr\'{e}d 1970. \textit{Probability Theory}.
Laszlo Vekerdi (trans.), Amsterdam: North-Holland.

\bibitem {ric:unify}Ricotta, Carlo and Laszlo Szeidl 2006. Towards a unifying
approach to diversity measures: Bridging the gap between the Shannon entropy
and Rao's quadratic index. \textit{Theoretical Population Biology}. 70: 237-43.

\bibitem {roseboom:abstract}Rozeboom, William W. 1968. The Theory of Abstract
Partials: An Introduction. \textit{Psychometrika} 33 (2 June): 133--67.

\bibitem {shannon:comm}Shannon, Claude E. 1948. A Mathematical Theory of
Communication. \textit{Bell System Technical Journal}. 27: 379-423; 623-56.

\bibitem {shannonweaver:comm}Shannon, Claude E. and Warren Weaver 1964.
\textit{The Mathematical Theory of Communication}. Urbana: University of
Illinois Press.

\bibitem {simp:md}Simpson, Edward Hugh 1949. Measurement of Diversity.
\textit{Nature}. 163: 688.

\bibitem {tamir-cohen:logicalentropy}Tamir, Boaz, and Eliahu Cohen. 2014.
Logical Entropy for Quantum States. \textit{ArXiv.org}. December. http://de.arxiv.org/abs/1412.0616v2.

\bibitem {tamir-cohen:hilbert-schmidt}Tamir, Boaz, and Eliahu Cohen. 2015. A
Holevo-Type Bound for a Hilbert Schmidt Distance Measure. \textit{Journal of
Quantum Information Science} 5: 127--33.

\bibitem {tsallis:entropy}Tsallis, Constantino 1988. Possible Generalization
for Boltzmann-Gibbs Statistics. \textit{J. Stat. Physics} 52: 479--87.

\bibitem {tsallis:nonext-sm}Tsallis, Constantino. 2009. \textit{Introduction
to Nonextensive Statistical Mechanics}. New York: Springer Science+Business Media.

\bibitem {uffink:phd}Uffink, Jos. 1990. \textit{Measures of Uncertainty and
the Uncertainty Principle }(PhD thesis). Utrecht Netherlands: University of Utrecht.

\bibitem {wilkins:merc}Wilkins, John 1707 (1641). \textit{Mercury or the
Secret and Swift Messenger}. London.

\bibitem {yeung:outlook}Yeung, Raymond W. 1991. A New Outlook on Shannon's
Information Measures. \textit{IEEE Trans. on Information Theor}y 37 (3): 466--74.

\bibitem {yeung:firstcourse}Yeung, Raymond W. 2002. \textit{A First Course in
Information Theory}. New York: Springer Science+Business Media.
\end{thebibliography}
\end{document}